%% file: main.tex
\documentclass{article}

\usepackage{microtype}      %
\usepackage{graphicx}
\usepackage{booktabs}       %
\graphicspath{{figures/}}

\usepackage{hyperref}

\usepackage[preprint]{icml2026}

\usepackage{amsmath}
\usepackage{amssymb}        %
\usepackage{mathtools}

\usepackage[capitalize,noabbrev]{cleveref}  %

\usepackage{xcolor}
\usepackage{tikz}
\usetikzlibrary{arrows.meta,decorations.pathreplacing}
\definecolor{okorange}{HTML}{E69F00}
\definecolor{okblue}{HTML}{0072B2}
\definecolor{okverm}{HTML}{D55E00}

\icmltitlerunning{Detecting AI Training in Datacentre Power}

\begin{document}

\twocolumn[
  \icmltitle{Detecting AI Training in Datacentre Power}

  \begin{icmlauthorlist}
    \icmlauthor{Tom Kimpson}{unimelb,mats}
    \icmlauthor{Mauricio Baker}{oxford}
    \icmlauthor{Emlyn Graham}{mats}
  \end{icmlauthorlist}

  \icmlaffiliation{unimelb}{School of Mathematics and Statistics, The University of Melbourne, Parkville, VIC 3010, Australia}
  \icmlaffiliation{oxford}{Department of Engineering Science, University of Oxford, Oxford, UK}
  \icmlaffiliation{mats}{Machine Alignment, Transparency \& Security (MATS)}

  \icmlcorrespondingauthor{Tom Kimpson}{tom.kimpson@unimelb.edu.au}

  \icmlkeywords{AI governance, compute verification, power side channel, cyclostationary detection}

  \vskip 0.3in
]

\printAffiliationsAndNotice{}

\begin{abstract}
AI governance agreements require methods to detect large model training runs on a monitored computing cluster, without trusting the operator. We develop a method to detect synchronous model training in facility-level power by Viterbi tracking of the wandering iteration cadence. On simulated datacentre power, when the operator makes no attempt to hide, the Viterbi tracker detects at least $96\%$ of training runs against a declared inference workload, at a $1\%$ false-alarm rate and within $0.04$ of the Neyman--Pearson optimum, given a meter that samples at $2$\,Hz or faster. Of eight adversarial strategies an operator might use to hide, only extreme variation of the work
per iteration evades the tracker. Its strongest form hides training in up to $73\%$ of five-minute
observations, but over a day at least $87\%$ of such runs are still caught, at a $0.1\%$ false-alarm rate per run. This adversarial
strategy costs no throughput, but in small-scale training the operator needs
$93$--$267\%$ more compute to reach the same loss. A modest passive meter is sufficient to make hidden training both detectable and expensive, which
makes power metering a promising part of compute verification that does not
depend on trusting the operator.

\end{abstract}

\section{Introduction}
\label{sec:intro}

AI governance agreements will require regulating training runs \citep{shavit2023chinchilla,sastry2024computing,baker2025verifying,heim2024thresholds}. This necessitates an external verifier to identify when training is occurring on a monitored cluster, against an operator who is trying to hide their training. There are multiple channels that could provide the verifier with diagnostic information to identify hidden training. These include on-chip mechanisms that log activities for later analysis \citep{shavit2023chinchilla,ogara2025hardware}, network taps that monitor data exchanged between chips, or human-centred approaches such as whistleblower programs or interviews of operator personnel \citep{baker2025verifying}. However, each of these channels requires trusting either hardware that is under the control of the operator, or the operator personnel themselves. Conversely, off-chip sensors that take analog, physical measurements of the cluster (e.g. power or electromagnetic emissions) do not require trusting the operator. Such sensors could be retrofitted to existing datacentres (e.g. clamping a power meter on an input cable) and secured against tampering by the operator. 

AI training has already been shown to leave a signature in the power drawn by the hardware that can serve as a diagnostic to identify and separate training from inference. A $10$\,MHz current probe clamped around the GPU's PCIe power supply cables can separate training from
inference and from non-AI work at $97\%$ accuracy \citep{gargiulo2026workload}, a result robust against many but not all adversarial strategies. On-chip $1$\,Hz telemetry counters that measure power (among other telemetry: utilisation, memory, etc.) can identify training at $98.2\%$ accuracy, falling to $43$--$87\%$ against the hardest workloads when they are adversarially disguised \citep{rahman2026telemetry}. In practice, any realistic AI governance agreements that use off-chip sensors will rarely have such fine-grained access to the GPU hardware. Power meters that can be retrofitted to existing facilities and secured against tampering are likely to sit at the level of the rack or facility input, with sampling frequencies ${\sim}10$\,Hz (i.e. six orders of magnitude slower than the board probe) and behind a power-delivery path that low-passes and dilutes the signal. At this level the training signal is still visible, and large enough that datacentre and grid operators treat it as an engineering control problem~\citep{patel2023polca,ko2025widearea}. 

At the facility meter, the signature of training is its iteration cadence. Each
iteration alternates a high-power compute phase with a low-power communication
phase, about once a second, and this cycle puts a narrow line in the power
spectrum. Detecting that line at a facility meter is harder than at the board,
for two reasons. First, the cadence of a real run is not fixed. Stragglers, degraded
hardware and temperature change the iteration time over seconds to hours
(\cref{subsec:ko}), so the spectral line wanders, and a test that looks for a
sharp peak at one frequency loses it. Second, an operator who wants to hide can
make the line wander on purpose, or smear it away, by inserting idle time
between iterations or by varying the work done in each. Hiding is not free,
however. A training operator cannot reshape its
signal freely; every distortion of the schedule costs throughput, learning
efficiency, or both. The question is therefore not whether training can be
hidden, but whether it can be hidden cheaply.

In this paper we develop a method to detect training in facility-level power
measurements against an operator who is trying to hide it, and we measure what
hiding costs. We model the power of a datacentre running training and inference
(\cref{sec:scenario}) and pass it through a model of an off-chip meter
(\cref{sec:threat}). To follow the wandering line we borrow a Viterbi line
tracker from passive sonar and continuous-gravitational-wave searches
(\cref{sec:methods}). Against an operator who does not hide, the tracker detects
training in almost every five-minute window, within $0.04$ of an optimal
ceiling, while fixed-frequency tests fail once the cadence drifts
(\cref{sec:results}). It needs only a meter that samples at $2$\,Hz or faster
and averages over no more than $0.5$\,s. We then test eight ways for an operator
to hide (\cref{sec:frontier}). Fixed-frequency tests are defeated cheaply:
varying the work per iteration hides $88\%$ of windows at no cost in throughput.
The tracker is defeated only by extreme work variation, and that has a cost in
learning instead: in small-scale training, the operator needs $93$--$267\%$ more
compute to reach the same loss. Over a day of observation, the strongest
strategy we found escapes detection in $13\%$ of runs, at a run-level
false-alarm rate of $0.1\%$ (\cref{subsec:runs}).

Our main conclusion is that a passive power meter outside the operator's control
can detect synchronous training, and that hiding from it is expensive. Two
limits apply. The meter detects a periodic schedule, not training itself, so
non-training loads with a training-like cycle also trip it
(\cref{subsec:identifiability}). And our evidence is synthetic: the traces come
from a model, not from a measured multi-GPU facility
(\cref{subsec:limitations}).

This paper is organised as follows. \Cref{sec:scenario} gives the model of
datacentre training and inference power that generates every trace below, and
\cref{sec:threat} fixes what the verifier sees through an off-chip meter and
what a concealing operator may do to hide from it. \Cref{sec:methods} develops the
detector, a Viterbi line tracker; \cref{sec:results} reports its power against
the inference null and the minimum meter it needs; and \cref{sec:frontier}
prices what it costs a concealing operator to defeat it. \Cref{sec:discussion}
asks how likely a concealing operator is to be detected over a run, covers the
non-training loads that imitate training, and sets out the limitations.
\Cref{sec:conclusion} concludes.

\section{A model of datacentre training power}
\label{sec:scenario}
In this section we build a model of the power draw from a datacentre running AI workloads. This model will serve as a synthetic data generator to test our methods, in lieu of having access to real empirical data. We adopt the stochastic model of AI training power from \citet{ko2025widearea}, developed for the analysis of grid stability, modified to add an Ornstein--Uhlenbeck drift on the training frequency. The reader is referred to \citet{ko2025widearea} for the full training model and its context.

This section is organised as follows. \Cref{subsec:ko} describes the power draw from a
training workload, following \citet{ko2025widearea}, and adds our drift of its cadence.
\Cref{subsec:finetune} describes the power draw from a fine-tuning workload. \Cref{subsec:null} describes the power draw from an inference workload. \Cref{subsec:aggregate} combines these workloads into the load of a whole
datacentre: one dominant workload, either training or inference, over a
background of smaller trainings and fine-tunings. \Cref{tab:params} collects
every parameter value of the model. 

\subsection{Training}
\label{subsec:ko}

Multi-GPU AI training of the kind that typifies frontier models is composed of a two-phase iterative cycle. First there is a power intensive, high compute ``up'' phase where the forward/backward pass happens. Second there is a low power, low compute, communication ``down'' phase where the gradient all reduce happens. 

Iteration $i$ has period
\begin{equation}
	T_i \;=\; \frac{1}{f_0\,(1+\xi_i)}, \qquad \xi_i \sim \mathcal{N}(0,\sigma_\xi^2),
	\label{eq:period}
\end{equation}
where $f_0$ is the baseline fluctuation frequency. For a given training run $f_0$ is fixed. Between training runs $f_0$ can vary depending on model size, batch size, parallelism layout, interconnect and hardware \citep{patel2024characterizing,jain2024pal}. To describe the inter-workload variation, we take $f_0 \sim \mathcal{U}(f^{\mathrm{lo}}, f^{\mathrm{hi}})$. The period splits into an up-phase of duration
$\rho_i T_i$ and a down-phase of duration $(1-\rho_i) T_i$, with the phase ratio
$\rho_i \sim \mathcal{U}(\rho^{\mathrm{lo}}, \rho^{\mathrm{hi}})$ redrawn each iteration. Writing $t$ for time
within iteration $i$, $P(t)$ for the power the run draws above the facility's static
overhead (added in \cref{subsec:obsmap}), and $P_{\mathrm{peak}}$ for its peak,
\begin{equation}
	\frac{P(t)}{P_{\mathrm{peak}}} =
	\begin{cases}
		1 + \Delta^{\mathrm{up}}_i + \nu^{\mathrm{up}}(t), & t < \rho_i T_i,\\[2pt]
		1 - \Delta^{\mathrm{dn}}_i + \nu^{\mathrm{dn}}(t), & t \geq \rho_i T_i,
	\end{cases}
	\label{eq:ko}
\end{equation}
where the per-phase level offsets are
$\Delta^{\mathrm{up}}_i \sim \mathcal{N}(0, \sigma_\Delta^2)$ and
$\Delta^{\mathrm{dn}}_i \sim \mathcal{N}(\mu_\Delta, \sigma_\Delta^2)$, and
$\nu^{\mathrm{up}}$, $\nu^{\mathrm{dn}}$ are zero-mean intra-phase Gaussian
fluctuations with standard deviations $\kappa_\nu s^{\mathrm{up}}_i$ and
$\kappa_\nu s^{\mathrm{dn}}_i$. The $s^{\mathrm{up}}_i$ and $s^{\mathrm{dn}}_i$
are redrawn each phase, uniformly from the ranges
$[s^{\mathrm{up,lo}}, s^{\mathrm{up,hi}}]$ and
$[s^{\mathrm{dn,lo}}, s^{\mathrm{dn,hi}}]$, and $\kappa_\nu$ is a fixed
scale factor. \Cref{tab:params} lists the values
of every parameter in this section.

\Cref{eq:period,eq:ko} are the exact training workload model from \citet{ko2025widearea}. We make one modification, adding a slow drift to the cadence $f_0$ itself. The jitter $\xi_i$ does not produce drift. Because it is redrawn independently
each iteration, the cadence averaged over many iterations is still $f_0$. The
timing errors accumulate as a random walk in phase, which broadens the spectral
line into a Lorentzian of full width $\approx 2\pi\sigma_\xi^2 f_0$ (to leading
order in $\sigma_\xi$) but leaves its centre at $f_0$. Real runs vary differently. Specifically, the iteration time changes in ways that persist.
Every training step waits for the slowest GPU, so one slow component slows the
whole run. In a production LLM cluster, these slowdowns last for the whole job
rather than varying from step to step \citep{lin2025stragglers}. Hardware can
also keep running but degrade: CPU contention, overheating GPUs and congested
network links cut throughput by tens of percent for periods of tens of seconds
to hours \citep{wu2024falcon}. Over a day,
temperature shifts throughput by a further $1$--$2\%$
\citep{grattafiori2024llama3}. 

We model this by letting the cadence of iteration $i$ be $f_0 g_i$, where $g_i$
is a discrete Ornstein--Uhlenbeck (first-order autoregressive) process that
starts at $g_0 = 1$ and reverts to $1$:
\begin{equation}
	g_i - 1 \;=\; (1-\theta)\,(g_{i-1} - 1) \;+\; \epsilon_i,
	\qquad \epsilon_i \sim \mathcal{N}(0, \sigma_g^2).
	\label{eq:ou}
\end{equation}
The period of \cref{eq:period} becomes
\begin{equation}
	T_i \;=\; \frac{1}{f_0\, g_i\,(1+\xi_i)},
	\label{eq:period_drift}
\end{equation}
which reduces to \cref{eq:period} when $g_i \equiv 1$. The stationary variance
of \cref{eq:ou} is $\sigma_g^2/(2\theta - \theta^2)$, so we set
\begin{equation}
	\sigma_g \;=\; \frac{\sigma_f}{f_0}\sqrt{2\theta - \theta^2},
	\label{eq:ou_sigma}
\end{equation}
which makes the stationary standard deviation of the instantaneous cadence
$f_0 g_i$ equal to $\sigma_f$, in Hz. The autocorrelation of $g_i$ at lag $k$
is $(1-\theta)^k$, so the cadence stays correlated over  $\sim 1/\theta$
iterations. Over that span it wanders by $\sigma_f$, which can exceed both the
line width set by $\xi_i$ and the frequency resolution of a trace. We clip
$g_i$ from below at $g_{\min}$ so the period stays finite. We hold $\theta$ fixed and treat $\sigma_f$ as a free
parameter, sweeping it in \cref{sec:results}.

\subsection{Fine-tuning}
\label{subsec:finetune}
Fine-tuning workloads can be represented using the same structure as in \cref{subsec:ko}, with different parameters (\cref{tab:params}). We denote the fine-tuning frequency as $f_1$. Compared with training, it has a lower cadence
band, a longer tail phase (larger $\rho_i$), a deeper tail-to-idle swing (larger
$\mu_\Delta$), and slightly smaller level offsets and intra-phase noise.
We apply the cadence drift of \cref{eq:ou} only to the dominant run;
fine-tuning and background training runs keep a fixed cadence ($\sigma_f = 0$).

\subsection{Inference}
\label{subsec:null}
\citet{ko2025widearea} do not present a model for power draw of inference workloads. We now construct one. We superpose three structures on a constant mean serving power load $B$. First, token generation (decode). Serving systems use continuous batching,
which adds and removes requests from the batch at every step
\citep{yu2022orca,kwon2023vllm}. A per-step cadence exists, but the batch
composition reshuffles every step. Unlike training, there is therefore no
characteristic frequency: decode power smears across a broad band with no
coherent frequency path. We model it as band-limited noise $b(t)$ of unit RMS
over $[f_b^{\mathrm{lo}}, f_b^{\mathrm{hi}}]$, scaled by an amplitude $A$. Second, expert-load imbalance. Under sparse mixture-of-experts routing, tokens
are not spread evenly across the experts \citep{shazeer2017moe}, so the work per
step varies with the mix of requests. We model this as a slow envelope on the
decode power: a unit-variance Ornstein--Uhlenbeck process $u(t)$ with
correlation time $\tau_u$, entering at depth $m$. Amplitude modulation spreads
the decode energy into sidebands, which makes inference harder to tell apart from training. Third, prompt processing (prefill). Each request begins with a prompt phase that
is brief, compute-bound, and draws more power than token generation
\citep{patel2024characterizing,patel2024splitwise}. We model these phases as box
pulses $s(\cdot)$ of duration $d_p$ and height $a$, arriving at times $\tau_k$
from a Poisson process of rate $\lambda_p$. These memoryless arrivals carry no
line and simply raise the broadband in-band level. Together,
\begin{equation}
	\frac{P^{\mathrm{inf}}(t)}{P_{\mathrm{peak}}} \;=\;
	B \;+\; A\,\bigl[1 + m\,u(t)\bigr]\,b(t) \;+\; \sum_k a\,s(t - \tau_k).
	\label{eq:null}
\end{equation}
The forms of these structures follow how LLM serving systems schedule work;
their amplitudes are our choice (\cref{tab:params}). We set them so that
inference is a loud confound rather than a quiet one. We model it as one
serving instance, whereas $M$ independent replicas would add incoherently and
so fluctuate only as $\sqrt{M}$ (see \cref{app:generator}). As
\cref{fig:scenario_traces} shows, it delivers more in-band power than training
but an order of magnitude less spectral contrast.

\subsection{Aggregate}
\label{subsec:aggregate}
A datacentre rarely runs one job at a time. Following \citet{ko2025widearea} we write the load behind a single meter as the aggregate of one dominant training run over a background of
smaller trainings and fine-tunings,
\begin{equation}
	P_{\mathrm{AI}}(t) \;=\; P^{\mathrm{main}}(t)
	\;+\; \sum_{j=1}^{N_{\mathrm{tr}}} P^{\mathrm{tr}}_j(t)
	\;+\; \sum_{k=1}^{N_{\mathrm{ft}}} P^{\mathrm{ft}}_k(t).
	\label{eq:aggregate}
\end{equation}
where $P^{\mathrm{main}}$ is the power of the dominant training run,
$P^{\mathrm{tr}}_j$ that of the $j$-th of $N_{\mathrm{tr}}$ smaller training
runs, and $P^{\mathrm{ft}}_k$ that of the $k$-th of $N_{\mathrm{ft}}$
fine-tuning runs. Each term is an independent draw of \cref{eq:ko}, with its own
cadence, phase ratios, and level offsets. $P^{\mathrm{main}}$ and the
$P^{\mathrm{tr}}_j$ use the training parameters and the $P^{\mathrm{ft}}_k$ use
the fine-tuning parameters; only $P^{\mathrm{main}}$ carries the cadence drift
of \cref{eq:ou}. Peak power is split between the dominant run, the small
trainings, and the fine-tunings in the ratio
$w_{\mathrm{main}} : w_{\mathrm{tr}} : w_{\mathrm{ft}}$, and each background
share is divided equally among its members. Each small training draws its cadence from the same
band as the dominant run, so the background carries weak lines of its own
inside the search band.

\begin{figure*}[t]
	\centering
	\includegraphics[width=\linewidth]{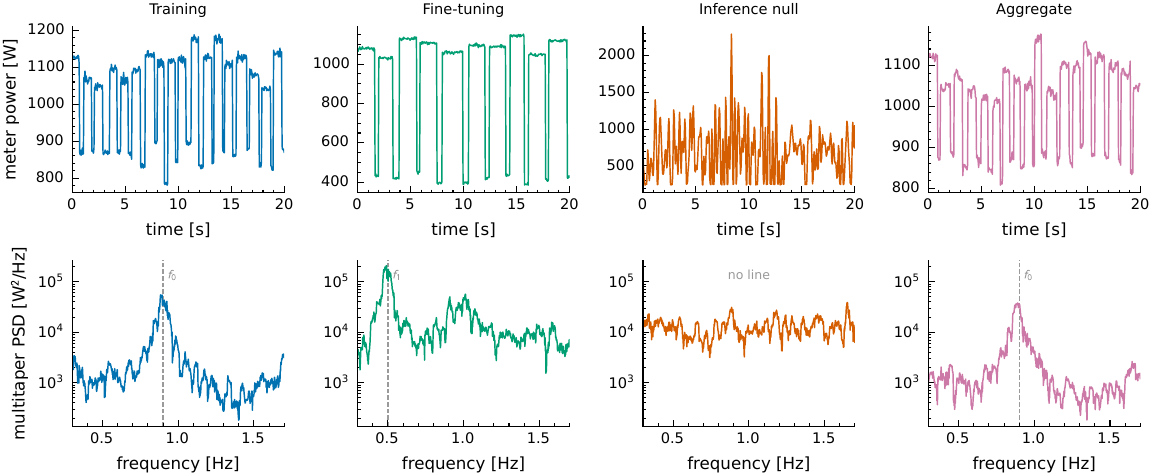}
	\caption{The four workload classes at the power meter, on the nominal channel
	(20\,Hz sampling, no filtering or decimation, $\sigma_\eta = 4$\,W). Top:
	a 20\,s excerpt. Bottom: the multitaper spectrum over the
	$0.3$--$1.7$\,Hz search band, with the true cadence marked. Training,
	fine-tuning, and the aggregate carry a line at $f_0$; the inference null
	does not. Cadences
	are pinned for legibility ($f_0 = 0.9$\,Hz training, $0.5$\,Hz
	fine-tuning); elsewhere they are drawn from the Ko bands.}
	\label{fig:scenario_traces}
\end{figure*}

We define the null aggregate as the same datacentre with the dominant training
run replaced by inference (\cref{subsec:null}). The background is unchanged,
\begin{equation}
	P^{\mathrm{null}}_{\mathrm{AI}}(t) \;=\; P^{\mathrm{inf}}(t)
	\;+\; \sum_{j=1}^{N_{\mathrm{tr}}} P^{\mathrm{tr}}_j(t)
	\;+\; \sum_{k=1}^{N_{\mathrm{ft}}} P^{\mathrm{ft}}_k(t),
	\label{eq:aggregate_null}
\end{equation}
with $P^{\mathrm{inf}}$ from \cref{eq:null} at the same
share $w_{\mathrm{main}}$. Training and inference therefore get the same share
of peak power, but inference draws less power on average. Both classes contain
the background's weak lines. \Cref{fig:scenario_traces} shows the four resulting workload classes as
a verifier would see them.

\begin{table}[t]
	\centering
	\footnotesize
	\setlength{\tabcolsep}{3pt}
	\begin{tabular}{@{}llcc@{}}
		\toprule
		& Symbol & Train & Fine-tune \\
		\midrule
		Cadence band [Hz] & $[f^{\mathrm{lo}}, f^{\mathrm{hi}}]$ & $[0.5, 1.5]$ & $[0.3, 0.7]$ \\
		Period jitter & $\sigma_\xi$ & $0.1$ & $0.1$ \\
		Phase ratio & $[\rho^{\mathrm{lo}}, \rho^{\mathrm{hi}}]$ & $[0.55, 0.8]$ & $[0.7, 0.9]$ \\
		Offset spread & $\sigma_\Delta$ & $0.05$ & $0.03$ \\
		Down-phase offset & $\mu_\Delta$ & $0.3$ & $0.8$ \\
		Up noise range & $s^{\mathrm{up}}$ & $[0.02, 0.05]$ & $[0.01, 0.03]$ \\
		Down noise range & $s^{\mathrm{dn}}$ & $[0.01, 0.03]$ & $[0.005, 0.02]$ \\
		Noise scale-down & $\kappa_\nu$ & $0.15$ & $0.15$ \\
		\midrule
		\multicolumn{4}{@{}l}{\emph{Cadence drift (dominant run only)}} \\
		Reversion [iter$^{-1}$] & $\theta$ & \multicolumn{2}{c}{$0.02$} \\
		Excursion std [Hz] & $\sigma_f$ & \multicolumn{2}{c}{$0$ (swept)} \\
		Drift floor & $g_{\min}$ & \multicolumn{2}{c}{$0.1$} \\
		\midrule
		\multicolumn{4}{@{}l}{\emph{Aggregate}} \\
		Background counts & $N_{\mathrm{tr}}, N_{\mathrm{ft}}$ & \multicolumn{2}{c}{$4, 4$} \\
		Power shares & $w_{\mathrm{main}}{:}w_{\mathrm{tr}}{:}w_{\mathrm{ft}}$ & \multicolumn{2}{c}{$9 : 0.5 : 0.5$} \\
		\midrule
		\multicolumn{4}{@{}l}{\emph{Inference null}} \\
		Mean serving load & $B$ & \multicolumn{2}{c}{$0.5$} \\
		Decode RMS & $A$ & \multicolumn{2}{c}{$0.40$} \\
		Decode band [Hz] & $[f_b^{\mathrm{lo}}, f_b^{\mathrm{hi}}]$ & \multicolumn{2}{c}{$[0.1, 4]$} \\
		MoE corr.\ time [s] & $\tau_u$ & \multicolumn{2}{c}{$5$} \\
		MoE depth & $m$ & \multicolumn{2}{c}{$0.4$} \\
		Prefill rate [Hz] & $\lambda_p$ & \multicolumn{2}{c}{$0.08$} \\
		Prefill height & $a$ & \multicolumn{2}{c}{$0.15$} \\
		Prefill duration [s] & $d_p$ & \multicolumn{2}{c}{$0.1$} \\
		\midrule
		\multicolumn{4}{@{}l}{\emph{Scale}} \\
		Peak power [W] & $P_{\mathrm{peak}}$ & \multicolumn{2}{c}{$850$} \\
		Static overhead [W] & $P_0$ & \multicolumn{2}{c}{$250$} \\
		\bottomrule
	\end{tabular}
	\caption{Workload-model parameters of \cref{sec:scenario}. Training and
	fine-tuning values are Ko and Zhu's \citep{ko2025widearea}, except
	$\kappa_\nu$, the drift parameters, and $\sigma_\xi$ for fine-tuning (Ko and
	Zhu list it for training only; we reuse it). The inference null is ours.
	Ko and Zhu's intra-phase noise is a sub-millisecond fluctuation, and each
	sample of our tens-of-hertz generator averages over many of them, so we
	scale its standard deviation by $\kappa_\nu$; the $s$ ranges are Ko and
	Zhu's values before scaling.
	Amplitudes are fractions of the peak power $P_{\mathrm{peak}}$. The drift
	excursion $\sigma_f$ is swept in \cref{sec:results}. The scale values are
	ours, and are those of a single accelerator board; only the workload's size
	relative to the meter's additive terms affects detection
	(\cref{subsec:channel}).}
	\label{tab:params}
\end{table}

\section{Measurement and threat model}
\label{sec:threat}

This section specifies how the power drawn by the datacentre reaches the verifier (i.e. the measurement model), and what the facility operator can control (i.e. the threat model). The verifier wants to know whether a monitored facility is running undeclared training. We frame this as a hypothesis test. Under the null hypothesis $\mathcal{H}_0$, the facility runs its declared workload, i.e. \cref{eq:aggregate_null}. Under the alternative hypothesis $\mathcal{H}_1$, the facility replaces inference with a training run, i.e. \cref{eq:aggregate}. \Cref{subsec:channel} outlines the channel through which either hypothesis reaches the verifier. \Cref{subsec:adversary} outlines who controls the load, what they know, and what
they may change. 

\subsection{The measurement model}
\label{subsec:channel}
\label{subsec:obsmap}

\begin{figure*}[t]
	\centering
	\input{fig_power_chain_v2}
	\caption{How the power drawn by the workloads reaches the verifier. The
	workloads of \cref{sec:scenario} and a static overhead sum to the facility
	power $P_{\mathrm{fac}}$, \cref{eq:fac}. This passes through the delivery
	path and the meter, and then additive terms are added, \cref{eq:meter}. The
	panels below show one training cadence at the four marked points as it is
	smoothed, sampled, and made noisy; they are illustrative. The shaded regions
	show which stages the operator and the verifier control
	(\cref{subsec:adversary}). Badges mark where the adversarial strategies of
	\cref{tab:attacks} act: schedule attacks and dilution (S) on the workload,
	and the meter strategy (M) on the channel. $\eta$ lumps instrument noise with
	interference that physically originates behind the meter, such as the
	power-controller line. In the nominal channel the delivery path
	passes the power through unchanged ($h_{\mathrm{del}} = \delta$) and the meter samples at $20$\,Hz with no integration
	(\cref{tab:meter}).}
	\label{fig:power_chain}
\end{figure*}
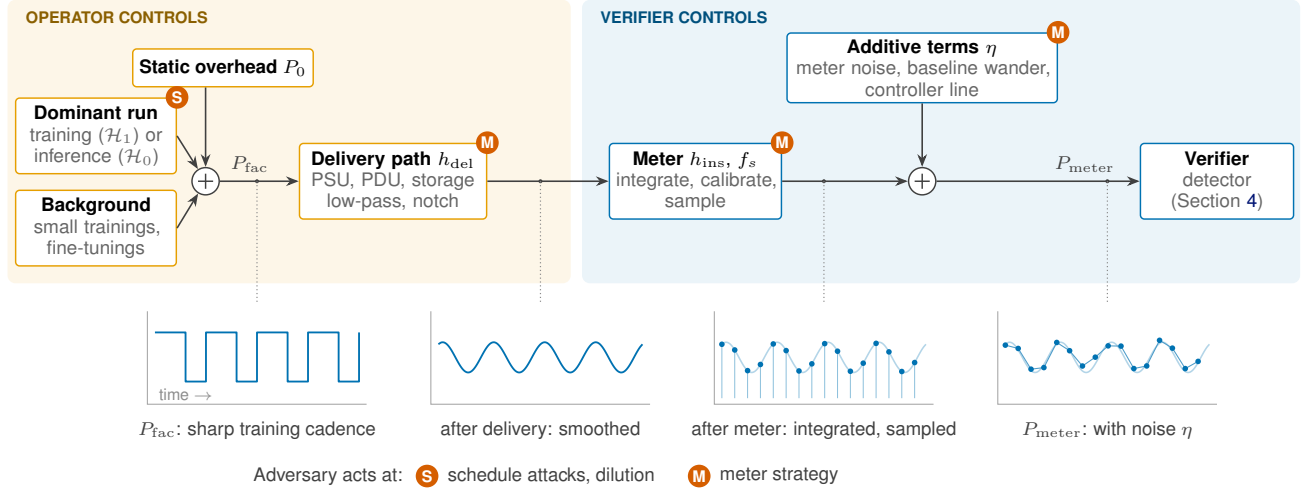

\Cref{sec:scenario} gives the power $P_{\mathrm{AI}}(t)$ drawn by the AI
workload above the facility's static overhead, in units of
$P_{\mathrm{peak}}$. The total power behind the meter is
\begin{equation}
	P_{\mathrm{fac}}(t) \;=\; P_0 + P_{\mathrm{AI}}(t),
	\label{eq:fac}
\end{equation}
with $P_{\mathrm{AI}}$ given by \cref{eq:aggregate} under $\mathcal{H}_1$, by
\cref{eq:aggregate_null} under $\mathcal{H}_0$, or by a single workload of
\cref{sec:scenario} when a result drops the background. The verifier holds one scalar time series from a meter outside the operator's
control. The meter reads
\begin{equation}
	\begin{gathered}
		P_{\mathrm{meter}}[n] \;=\; (h * P_{\mathrm{fac}})(t_n) + \eta[n],\\
		t_n = n/f_s, \qquad h \;=\; h_{\mathrm{ins}} * h_{\mathrm{del}}.
	\end{gathered}
	\label{eq:meter}
\end{equation}
The meter filters the facility power, samples it at rate $f_s$, and adds
interference and measurement noise $\eta[n]$ on its own sample grid. Round
brackets denote continuous time and square brackets index the meter's
discrete samples. The filter is a linear time-invariant system with impulse response $h$, which has
two parts: the delivery path $h_{\mathrm{del}}$ and the instrument
$h_{\mathrm{ins}}$. \Cref{fig:power_chain} traces the power from the workloads
to the verifier, and \cref{tab:meter} lists every stage of the map, with its
nominal setting and the range we sweep. The nominal channel, used throughout
unless stated, passes the facility power through unchanged ($h = \delta$),
samples it at $f_s = 20$\,Hz, and adds white meter noise of
$\sigma_\eta = 4$\,W. Each trace lasts $T = 300$\,s, a frequency resolution of
$1/T \approx 3$\,mHz. We take the three parts of \cref{eq:meter} in turn.

\begin{figure*}[t]
	\centering
	\includegraphics[width=\linewidth]{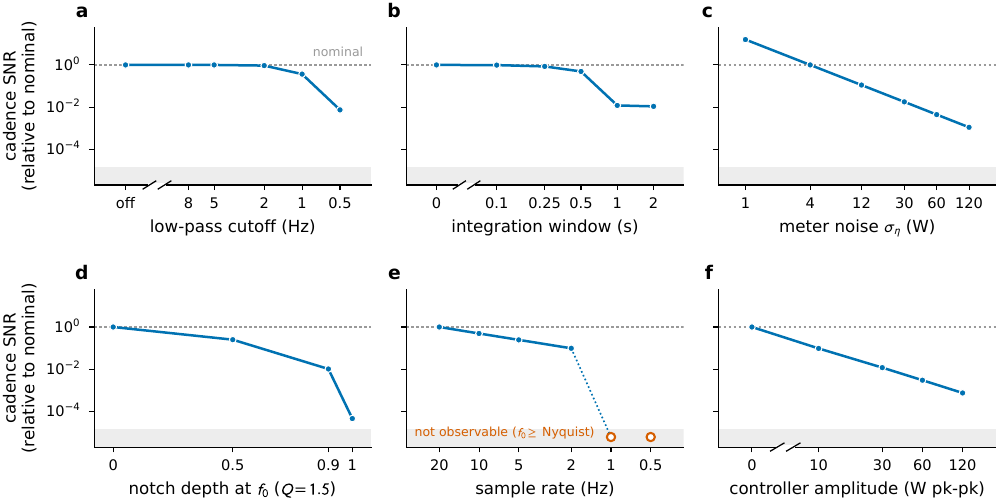}
	\caption{Sensitivity of the cadence signature to the observation map, on a fixed honest training workload ($f_0 = 0.9$\,Hz, median
	of 20 realisations; the interquartile range is narrower than the markers).
	Columns follow the stages of \cref{fig:power_chain}: the delivery path
	(a, d), the meter (b, e), and the additive terms (c, f). The measure is an
	effective signal-to-noise ratio: spectral power at the known $f_0$ over the
	in-band power the meter itself contributes, normalised to the nominal
	channel (dashed line). Attenuating axes lower the signal and additive ones
	raise the noise, so every panel reads the same way: down is worse for the
	verifier.
	Harsher channels lie to the right. Open circles mark sample rates at which
	$f_0$ reaches the Nyquist frequency, so the cadence is not observable at
	all. Note $f_0$ is pinned away from
	$1.0$\,Hz, which would coincide with both the $2$\,Hz Nyquist frequency and
	the first sinc null of a $1$\,s boxcar.}
	\label{fig:scenario_meter}
\end{figure*}

The delivery path $h_{\mathrm{del}}$ is everything between the accelerators
and the meter: power supplies, distribution, and any conversion or storage.
Following \citet{ko2025widearea} we assume it passes the power through
unchanged, such that $h_{\mathrm{del}}$ is a delta function and
$h_{\mathrm{del}} * P_{\mathrm{fac}} = P_{\mathrm{fac}}$. The low-pass and
notch filters of \cref{tab:meter} relax this assumption:
\cref{fig:scenario_meter} shows how much each attenuates the cadence, and
\cref{subsec:meter_spec} tests the notch filter against detection. The notch
filter represents an anti-resonance, i.e. a narrow band of frequencies that the
delivery network suppresses. If that band sits near the cadence, it removes the
training signal while leaving the rest of the spectrum untouched.

The instrument $h_{\mathrm{ins}}$ is the meter's own filtering: its
integration window and calibration gain. We model it as a calibration gain $g$ followed by a
trailing boxcar, so that each reading is the mean power over the preceding
window $\tau$: $h_{\mathrm{ins}}(t) = g/\tau$ for $0 \le t < \tau$ and zero
otherwise. Its frequency response has magnitude
\begin{equation}
	\lvert H_{\mathrm{ins}}(f) \rvert \;=\; g\,
	\left\lvert \frac{\sin(\pi f \tau)}{\pi f \tau} \right\rvert,
	\label{eq:boxcar}
\end{equation}
which passes slow variations and attenuates fast ones, with a zero at every
multiple of $1/\tau$. A window longer than about $0.6$\,s puts the first zero
inside the $0.3$--$1.7$\,Hz search band, and a $1$\,s window puts it at the
band centre. In the nominal channel $g = 1$ and $\tau \to 0$, so
$h_{\mathrm{ins}}$ is also a delta function, and the meter samples at
$f_s = 20$\,Hz. Unlike the delivery path, the instrument is the verifier's to
choose, and \cref{subsec:meter_spec} asks what $\tau$ and $f_s$ must be for
detection to hold.

The additive term $\eta[n]$ collects everything that reaches the readings
without passing through $h$ (\cref{tab:meter}). In the nominal channel it is
white meter noise of $\sigma_\eta = 4$\,W. Two further terms are available: a
slowly wandering baseline, standing for a drifting operating point such as
cooling load, and a controller limit cycle, a square wave whose frequency
itself wanders, modelled on the power-management oscillation measured on an
A100 (\cref{subsec:identifiability}). The controller is the first appearance of an
instrument line: a persistent, slowly wandering spectral line that the
facility's hardware produces with no training present.
\Cref{subsec:identifiability} returns to it.

Three numbers carry units. The peak power $P_{\mathrm{peak}} = 850$\,W and
overhead $P_0 = 250$\,W (\cref{tab:params}) are those of a single accelerator
board, and the meter noise $\sigma_\eta = 4$\,W is an illustrative choice,
about $0.5\%$ of the mean reading, not a measured value. None of the three is
critical. $P_0$ is constant, so it lies at zero frequency, outside the search
band. What sets detectability is the training line against everything else in
the band, and at the nominal setting that is dominated by the workloads
themselves; in the search band the inference null's power spectral density
exceeds that of the meter noise by about four orders of magnitude, and the
training trace's own off-line floor exceeds it by about three. Meter noise
binds only when it is raised far above $4$\,W, which
\cref{fig:scenario_meter} does across two orders of magnitude.

Two limits apply. The linear time-invariant form of $h$ is a first
approximation; a real delivery path may be operating-point dependent. And the
model quantises only in time, through the integration window and the sample
grid; it has no amplitude quantisation. Until simultaneous on-device and
external-meter measurement is performed, claims about what reaches an
off-chip sensor are predictions (\cref{subsec:limitations}).

\begin{table*}[t]
	\centering
	\footnotesize
	\begin{tabular}{@{}llccl@{}}
		\toprule
		Stage & Parameter & Nominal & Swept & Stands for \\
		\midrule
		\multicolumn{5}{@{}l}{\emph{Delivery path ($h_{\mathrm{del}}$; the identity in the workload model)}} \\
		Butterworth low-pass (order 2) & cutoff [Hz] & off & $8$--$0.5$ & PSU/PDU bandwidth, smoothing \\
		In-band notch & depth; centre [Hz]; $Q$ & off & $0.5$--$1$; $0.6$--$1.2$; $1.5$ & delivery-path anti-resonance \\
		\midrule
		\multicolumn{5}{@{}l}{\emph{Instrument ($h_{\mathrm{ins}}$ and sampling; chosen by the verifier)}} \\
		Trailing boxcar integration & window $\tau$ [s] & $0$ & $0.1$--$2$ & meter averaging \\
		Gain & $g$ & $1$ & --- & calibration factor \\
		Sampling (no anti-alias filter) & $f_s$ [Hz] & $20$ & $10$--$0.5$ & meter reporting rate \\
		\midrule
		\multicolumn{5}{@{}l}{\emph{Additive terms ($\eta$)}} \\
		Meter noise (white or AR(1)) & $\sigma_\eta$ [W]; $\tau_\eta$ [s] & $4$; white & $1$--$120$; $2$ & instrument noise \\
		Baseline wander (OU) & std [W]; corr.\ time [s] & off & $20$; $60$ & drifting operating point \\
		Controller limit cycle & amplitude [W pk-pk] & off & $10$--$120$ at $0.38$\,Hz & power-management loop \\
		\bottomrule
	\end{tabular}
	\caption{Stages of the observation map (\cref{eq:meter}). Nominal values
	define the channel used everywhere unless stated. The swept ranges are those
	of \cref{fig:scenario_meter} and the minimum-meter sweep of
	\cref{subsec:meter_spec}; the hostile channels of the meter strategy
	(\cref{tab:attacks}) are drawn from the same ranges. Every stage off except
	$\eta$ is the nominal channel; every stage off including $\eta$ is the
	identity.}
	\label{tab:meter}
\end{table*}

\Cref{fig:scenario_meter} shows how the channel attenuates the training
line. For a fixed training workload we vary one axis of \cref{tab:meter} at a
time and measure an effective signal-to-noise ratio at the cadence: the
spectral power at $f_0$ over the in-band power the meter itself adds,
relative to the nominal channel. Every axis lowers it. Within the range a real
meter might occupy, a $1$\,s integration window, whose first zero falls at
$1$\,Hz (\cref{eq:boxcar}), cuts it by about two orders of magnitude, and a
$1$\,Hz sample rate removes the line from the band entirely, because $f_0$ then lies above
the Nyquist frequency and aliases to $0.1$\,Hz. \Cref{app:generator} gives the
measure as an equation. This figure measures the signal that reaches the verifier, not how well any
detector uses the signal; \Cref{subsec:meter_spec} tests the detectors on the surviving signal and discusses the minimum meter specification.

\subsection{The threat model}
\label{subsec:adversary}

The verifier installs and owns the power meter.
The meter is tamper-evident, the verifier chooses its specification, and the
operator cannot forge the meter noise $\eta$. The operator controls everything behind the meter, e.g. which jobs run, how they
are scheduled, what else shares the facility, and the power-delivery path
itself. Energy storage or
power smoothing between the accelerators and the meter can low-pass or notch
the load, and datacentre and grid operators already treat smoothing training
load as an engineering problem \citep{ko2025widearea}. We represent partial
smoothing by the filtering stages of \cref{tab:meter}, and
\cref{subsec:meter_spec} asks how much of it the verifier can tolerate. A
store large enough to flatten the load entirely would erase any passive
signature; we do not model it, and discuss it among the limitations
(\cref{subsec:limitations}). The operator knows the
detector, its parameters, its threshold, and the meter specification. It does
not know the realisation of $\eta$. %

We consider three types of facility operator. A \emph{compliant} operator runs only its declared inference. A \emph{benign} operator trains without regard to being caught by the power meter.  A \emph{concealing} operator trains, knows the detector, and reshapes its workload to try to evade it. The concealing operator's objective is to train
without being detected by the verifier. They use different adversarial strategies to try to subvert the verifier's detection. \Cref{tab:attacks} lists the adversarial strategies considered in this work. They
fall into five categories: disturbing the timing of iterations, relocating the
cadence, reshaping the power waveform within an iteration, mixing the run with
other load, and degrading the channel. In principle the strategies may be composed, but we treat them individually. Adversarial strategies have a cost to the operator in terms of training throughput. Each strategy therefore trades concealment against cost. 

\begin{table*}[t]
	\centering
	\footnotesize
	\begin{tabular}{@{}p{2.3cm}p{7.4cm}p{4.3cm}c@{}}
		\toprule
		Strategy & What the operator changes & Budget knob (range) & Cost anchor \\
		\midrule
		\multicolumn{4}{@{}l}{\emph{Timing}} \\
		Period jitter & i.i.d.\ period jitter $\sigma_\xi$ of \cref{eq:period}, by inserting idle time & $\sigma_\xi \in [0.05, 0.7]$ & measured \\
		Cadence drift & OU excursion $\sigma_f$ of \cref{eq:ou} & $\sigma_f \in [0, 1.5]$\,Hz & measured \\
		Work variation & micro-steps per iteration $G_i$, drawn about $G_0 = 8$ with fractional spread $\sigma_G$; down-phase unchanged & $\sigma_G \in [0.05, 0.7]$ & measured \\
		Phase slip & Brownian slip of iteration boundaries at fixed mean period & slip std $\sigma_s \in [0.01, 0.2]$ periods & --- \\
		\midrule
		\multicolumn{4}{@{}l}{\emph{Frequency}} \\
		Relocation & cadence $f_0$ moved toward the search-band edges & $f_0 \in [0.35, 1.65]$\,Hz & --- \\
		\midrule
		\multicolumn{4}{@{}l}{\emph{Waveform}} \\
		Harmonic smoothing & raised-cosine ramp on phase transitions; comb dies, fundamental survives & ramp width $\in [0.1, 1]$\,s & --- \\
		Shape filling & down-phase level raised toward the up-phase at matched mean power & fill fraction $\phi \in [0, 1]$ & measured \\
		\midrule
		\multicolumn{4}{@{}l}{\emph{Mixing}} \\
		Dilution & training's share of the aggregate (\cref{eq:aggregate}) lowered & share $\in [0.2, 0.9]$ & --- \\
		\midrule
		\multicolumn{4}{@{}l}{\emph{Channel}} \\
		Meter & hostile $h$ and $\eta$: notch at the cadence; $1$\,s/$1$\,Hz integrating sampler; in-band controller; heavy coloured noise & four named channels & --- \\
		\bottomrule
	\end{tabular}
	\caption{Adversarial strategies available to the concealing operator. Each is
	implemented inside the generator of \cref{sec:scenario} and swept singly over
	the stated budget range in \cref{sec:frontier}. ``Measured'' marks strategies
	with a throughput-overhead anchor from the single-A100 campaign;
	the others were not run on hardware and are
	reported without a cost (\cref{sec:frontier}). \Cref{app:generator} defines each strategy's generator
change. The meter strategy
	degrades the channel rather than the schedule and is read as a channel
	requirement (\cref{subsec:meter_spec}), not priced.}
	\label{tab:attacks}
\end{table*}

Three things lie outside this threat model. First, the
operator does not tamper with the meter or forge its noise. Second, it does not run a
semantic decoy, that is, a load that reproduces the physical schedule of
training without training,  since manufacturing evidence of the activity it
is hiding buys a concealing operator nothing. Finally, it does not
move the training cadence outside the searched band or abandon synchronous iteration
altogether; both are scope restrictions on the population we target
(\cref{subsec:limitations}). Moving the cadence below the band is not free. To do it, the operator must
accumulate more micro-steps per step, which enlarges the batch and, past the
critical batch size, costs learning efficiency (\cref{sec:frontier}). The
verifier can also extend the band downward, at the price of longer spectrogram
frames.

\section{Detection method}
\label{sec:methods}

In this section we present a detection method that ingests  $P_{\mathrm{meter}}[n]$, \cref{eq:meter}, and decides between $\mathcal{H}_0$ and $\mathcal{H}_1$ (\cref{sec:threat}). The two hypotheses are separated by the iteration cadence of synchronous training. Under $\mathcal{H}_1$ the two-phase cycle of the dominant run (\cref{eq:ko}) puts a strong, narrow spectral line at its cadence $f_0$. Under $\mathcal{H}_0$ the inference that replaces it spreads its power across the band with no line
(\cref{subsec:null}), and the only lines left are the background's weak ones
(\cref{subsec:aggregate}). A line at a fixed frequency has classical tests. But
the cadence of a real run wanders by $\sigma_f$ even without an adversary
(\cref{eq:ou}), and a concealing operator can make it wander further or vary
the work per iteration, which defeats a fixed-frequency test cheaply
(see \cref{sec:frontier}). The detector therefore has to follow the line as it
moves.

This section is organised as follows. \Cref{subsec:viterbi} defines the
detector, a Viterbi line tracker borrowed from passive sonar and
continuous-gravitational-wave searches. \Cref{subsec:threshold} sets the threshold at which the detector declares training.
\Cref{subsec:comparators} defines the two fixed-frequency tests we compare it
against.

\subsection{Viterbi line tracking}
\label{subsec:viterbi}

We follow the line with the hidden-Markov frequency tracking developed for
passive sonar \citep{streit1990frequency} and carried into
continuous-gravitational-wave searches
\citep{suvorova2016hmm,bayley2019soap,kimpson2024anc}. The tracker has
three parts: a spectrogram of the trace, a contrast map built from it, and a
search for the best slowly varying path through that map.

Let $x[n] = P_{\mathrm{meter}}[n] - \bar
P_{\mathrm{meter}}$ be the mean-removed meter trace, $N = f_s T$ samples long.
We cut it into frames of $M$ samples that overlap by half, a hop of $H = M/2$,
and take the power spectrum of each:
\begin{equation}
	S_{kj} \;=\; \Bigl\lvert \sum_{m=0}^{M-1} w[m]\,x[jH + m]\,
	e^{-2\pi i k m / M} \Bigr\rvert^2,
	\label{eq:spectrogram}
\end{equation}
for frames $j = 0, \dots, N_t - 1$, where $w$ is a tapered (Tukey) window and $N_t = \lfloor (N - M)/H \rfloor + 1$
is the number of frames. Bin $k$ sits at frequency $f_k = k f_s / M$. We
choose $M$ so that each frame lasts $M/f_s = 16$\,s, which gives bins
$\Delta f = 1/16 = 62.5$\,mHz wide, fine enough to resolve a line near $1$\,Hz,
while leaving many frames to track over: a $300$\,s trace gives $N_t = 36$
frames, $8$\,s apart. Only the bins with $f_k \in \mathcal{B}$ below the
Nyquist frequency $f_s/2$ enter the tracker, $23$ of them in the nominal
channel (\cref{app:generator} covers low sample rates). Every detector in this paper searches the band
$\mathcal{B} = [0.3, 1.7]$\,Hz. This is the training cadence band
$[f^{\mathrm{lo}}, f^{\mathrm{hi}}] = [0.5, 1.5]$\,Hz of \cref{tab:params},
widened by $0.2$\,Hz on each side so that a cadence drawn near either edge
stays in the band as it wanders.

We normalise each frame by its median over the band,
\begin{equation}
	L_{kj} \;=\; \ln S_{kj} \;-\; \ln \tilde S_j,
	\label{eq:contrast}
\end{equation}
where $\tilde S_j$ is the median of $S_{kj}$ over the in-band bins of frame
$j$. $L_{kj}$ measures how far bin $k$ stands above the typical in-band level of
its own frame. The normalisation keeps a slow change in overall power from masquerading as a line.

\begin{figure*}[t]
	\centering
	\includegraphics[width=\linewidth]{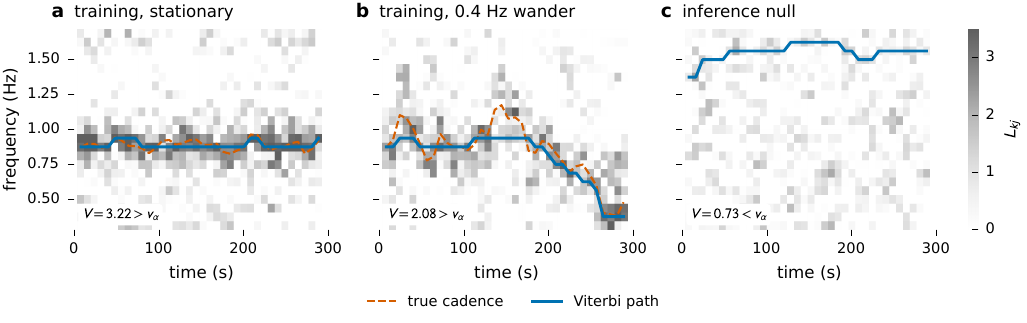}
	\caption{The Viterbi tracker on three $300$\,s traces at the nominal channel.
	Each panel is the contrast map $L_{kj}$ of \cref{eq:contrast} on the
	tracker's own grid ($16$\,s frames at half overlap, $62.5$\,mHz bins over
	$\mathcal{B}$), with the best path of \cref{eq:viterbi} in white and its
	score $V$ inset. Dashed: the true cadence, $1/$(median iteration period)
	in each frame, from the generator's iteration boundaries. Left: a
	stationary training cadence ($f_0 = 0.9$\,Hz), which the path sits on.
	Centre: the same run with $0.4$\,Hz of cadence wander, which the path
	follows across most of the band; it cuts the corner off the brief
	excursion near $150$\,s, where following the line costs more in jumps than
	it gains in contrast. Right: the inference null. The tracker still returns
	a best path, as it does for any input, but here it only strings together
	noise fluctuations and scores $V = 0.73$, below the threshold
	$v_\alpha \approx 0.91$ at $\alpha = 10^{-2}$ (\cref{subsec:threshold}).}
	\label{fig:viterbi_track}
\end{figure*}

The hidden state is the frequency bin occupied by the
cadence in each frame. The prior that a real cadence wanders slowly enters as
a penalty $\lambda$ per bin of frequency change between consecutive frames. The
detection statistic is the score of the best path, averaged over frames,
\begin{equation}
	V \;=\; \frac{1}{N_t}\,\max_{k_0,\dots,k_{N_t-1}}
	\Biggl[\,\sum_{j=0}^{N_t-1} L_{k_j j}
	\;-\; \lambda \sum_{j=1}^{N_t-1} \lvert k_j - k_{j-1}\rvert\Biggr],
	\label{eq:viterbi}
\end{equation}
with $\lambda = 1$ throughout. A path pays one unit of log-contrast for each
$62.5$\,mHz step it takes between frames, so it can follow a drift of about
one bin per frame (about $8$\,mHz\,s$^{-1}$) if the line it follows is bright
enough to pay for it. Dynamic programming evaluates the maximum over all
$23^{N_t}$ paths exactly, in time linear in $N_t$. A drifting line keeps a high
score because the path bends to follow it. Broadband power, such as the
continuous-batching floor of the inference null, does not, because no single
slowly varying path stays on top of it. \Cref{fig:viterbi_track} shows both cases on
the contrast map. Averaging over frames makes $V$
comparable across traces of different length. Note that $V$ tests whether a
persistent, slowly wandering line is present, not what produced it. Accordingly, strong lines that do not have a training provenance may also trigger the detector; this is addressed in
\cref{subsec:identifiability}.

\subsection{Setting the threshold}
\label{subsec:threshold}

The tracker declares training when $V$ exceeds a threshold $v_\alpha$, chosen
so that the false-alarm rate is $\alpha$:
\begin{equation}
	\Pr\bigl(V > v_\alpha \mid \mathcal{H}_0\bigr) \;=\; \alpha.
	\label{eq:far}
\end{equation}
We use $\alpha = 10^{-2}$ unless stated. $V$ has no analytic null
distribution. We therefore set $v_\alpha$ empirically, as the
$(1-\alpha)$ quantile of $V$ over simulated traces drawn under
$\mathcal{H}_0$, and measure detection rate as the fraction of $\mathcal{H}_1$
traces above it. The baseline methods of \cref{subsec:comparators} are thresholded
the same way.

In \cref{sec:results} the $\mathcal{H}_0$ population is the null aggregate of
\cref{eq:aggregate_null}. Its background carries the weak lines of the small
trainings and fine-tunings, so the threshold already accounts for any line
that is not the dominant run's. Calibrating on inference alone, with no
background, changes the tracker's detection rates by at most $0.02$
(\cref{app:ceiling}).

This calibrates $V$ against the modelled inference null. A deployed verifier
has to obtain the same threshold from data it holds, either from background traces
known to be free of training, or from
surrogates generated from each observed trace.

\subsection{Baseline methods}
\label{subsec:comparators}

Two simple fixed-frequency tests serve as baselines to compare our Viterbi method against. The first is a
spectral matched filter, which uses the peak of the power spectral density in the band,
relative to its median over all frequencies,
\begin{equation}
	D_{\mathrm{MF}} \;=\;
	\frac{\max_{f \in \mathcal{B}} \hat S(f)}{\operatorname{median}_f \hat S(f)},
	\label{eq:matched}
\end{equation}
where $\hat S$ is a Welch estimate of the spectrum of $x[n]$ from half-overlapping
Hann-windowed segments of $1024$ samples ($51.2$\,s at $f_s = 20$\,Hz). The
long segments concentrate a stationary line into one bin, which is the
test's strength; a wandering line smears across many bins and the peak
collapses. The second is the Thomson multitaper harmonic F-test
\citep{thomson1982multitaper,percival1993spectral}, maximised over the band,
\begin{equation}
	D_{\mathrm{F}} \;=\; \max_{f \in \mathcal{B}} F(f),
	\label{eq:ftest}
\end{equation}
where, at any fixed $f$ under a stationary, locally smooth null, $F(f) \sim
F(2, 2K-2)$, with $K = 7$ tapers of time--bandwidth product $NW = 4$. $F(f)$ is the ratio of
the power in a line at $f$ to the power left around it
(\cref{eq:thomson_f}, \cref{app:baselines}). The F-test needs a line that
stays phase-coherent over the whole trace. Even without drift, the period
jitter of \cref{eq:period} broadens the line to $\approx 2\pi\sigma_\xi^2 f_0
\approx 63$\,mHz at $1$\,Hz (\cref{subsec:ko}), wider than the $2NW/T \approx
27$\,mHz bandwidth of the tapers, so we expect the F-test to be weak at every
drift.

\section{Detection results}
\label{sec:results}

In this section we test the Viterbi tracker of \cref{sec:methods} on synthetic
meter traces. We assess if the Viterbi tracker can detect the training frequency, and what sampling settings are required by the power meter for the Viterbi tracker to work. We treat the operator as benign; an operator who reshapes the schedule
to hide the line is the subject of \cref{sec:frontier}. Under $\mathcal{H}_1$ the meter sees a dominant training run over
$N_{\mathrm{tr}} = 4$ small trainings and $N_{\mathrm{ft}} = 4$ fine-tunings
(\cref{eq:aggregate}). Under $\mathcal{H}_0$ it sees the null aggregate, with
inference in the dominant slot over the same kind of background
(\cref{eq:aggregate_null}). The shares are the defaults of \cref{tab:params},
$9 : 0.5 : 0.5$. Both classes therefore also carry the background's weak lines, and
the tracker has to find the dominant one among them.

This section is organised as follows. \Cref{subsec:power} measures the
tracker's detection rate as the cadence drifts, and compares it with the
fixed-frequency tests of \cref{subsec:comparators} and with an optimality
ceiling. \Cref{subsec:meter_spec} sweeps the meter and states the minimum
specification under which detection holds.

\subsection{Detection rate under cadence wander}
\label{subsec:power}

\begin{figure}[t]
	\centering
	\includegraphics[width=\columnwidth]{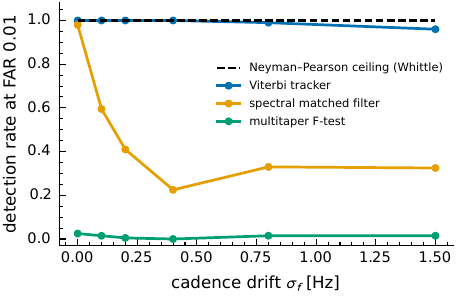}
	\caption{Detection rate at empirical false-alarm rate $10^{-2}$ against the
	null aggregate, versus cadence drift $\sigma_f$ of the dominant training run ($n=200$
	per class, $300$~s traces, shares $9 : 0.5 : 0.5$).
	The Viterbi tracker stays within $0.04$ of the Whittle Neyman--Pearson
	ceiling across the wander axis; the spectral matched filter collapses once the line moves and
	the multitaper F-test has no usable power. The ceiling is described in
	\cref{app:ceiling}.}
	\label{fig:drift_ceiling}
\end{figure}

We consider a range of cadence-drift  $\sigma_f \in \{0, 0.1, 0.2, 0.4, 0.8, 1.5\}$~Hz
(\cref{eq:ou_sigma}) in the dominant training run, with $n=200$ traces of
$300$~s per class and thresholds set at an empirical false-alarm rate of
$10^{-2}$. We benchmark the Viterbi method against a Neyman--Pearson optimal ceiling
(\cref{app:ceiling}). \Cref{fig:drift_ceiling} shows the result. The Viterbi tracker stays within $0.04$ of the optimal ceiling across the
whole $\sigma_f$ axis, with detection rate $1.00$ up to $0.4$~Hz of drift, $0.99$ at
$0.8$~Hz and $0.96$ at $1.5$~Hz. Conversely, the fixed-frequency baselines struggle. The spectral matched
filter detects a stationary line ($0.98$ at zero drift) but falls to $0.59$ at
$0.1$~Hz and $0.23$ at $0.4$~Hz, and the multitaper F-test has essentially no
power at empirical thresholds at any drift (${\leq}\,0.03$). 

\subsection{A minimum meter specification}
\label{subsec:meter_spec}

The results of \cref{subsec:power} assume the nominal meter of
\cref{tab:meter}: a sample rate of $20$~Hz, no integration window ($\tau \to 0$), no filtering
on the delivery path, and white meter noise of $\sigma_\eta = 4$~W. A real meter
may sample more slowly, average over a longer window, or sit behind a delivery
path that suppresses the cadence band. The meter is the verifier's choice, so it
is of interest to find the minimum specification it must meet for detection to
hold. We sweep the sample rate over $f_s \in \{20, 10, 5, 2, 1, 0.5\}$~Hz and
the integration window over $\tau \in \{0, 0.1, 0.25, 0.5, 1, 2\}$~s
(\cref{eq:boxcar}). We also sweep a notch in the delivery path, i.e. a band near the
cadence in which the delivery path suppresses the power swing
(\cref{subsec:obsmap}). Its centre ranges over $0.6$--$1.2$~Hz and its depth,
the fraction of the signal it removes at the centre, over $50$--$100\%$, with
the meter otherwise nominal. Throughout, the honest aggregate is scored against the null
aggregate, with thresholds at an empirical false-alarm rate of $10^{-2}$ as in
\cref{subsec:power}.

\begin{figure}[t]
	\centering
	\includegraphics[width=\columnwidth]{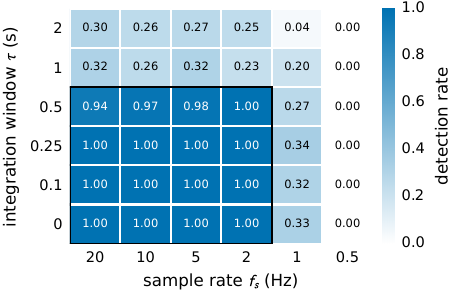}
	\caption{Minimum meter specification: Viterbi detection rate at empirical
	false-alarm rate $10^{-2}$ over sample rate $f_s$ $\times$ integration window $\tau$, the
	honest aggregate against the null aggregate ($n=200$ per class). The live region is
	$f_s \ge 2$~Hz and $\tau \le 0.5$~s.}
	\label{fig:meter_boundary_viterbi}
\end{figure}

The Viterbi tracker holds full power (detection
rate ${\gtrsim}0.9$) if and only if the meter satisfies three
conditions together (\cref{fig:meter_boundary_viterbi}), one on each part of
\cref{eq:meter}. First, on the sampling, a rate of at least about $2$~Hz: rates of $20$, $10$, $5$, and $2$~Hz all hold the Viterbi tracker
near $1.0$, $1$~Hz collapses it to $0.33$ as the Nyquist edge crosses the band,
and $0.5$~Hz is dead (every detector at chance). At $2$~Hz the Nyquist frequency is $1$~Hz, so half the cadence band
($1$--$1.5$~Hz) lies above it. These cadences fold back into the band, because our simulated meter has no
anti-alias filter, and the tracker finds the folded line. A meter with a sharp
anti-alias filter would remove these cadences instead, and would need
$f_s \gtrsim 3.4$~Hz to keep the whole search band. Second, on the instrument's
impulse response, an integration window $\tau \lesssim 0.5$~s: at $20$~Hz
sampling $\tau = 0.5$~s still holds ($0.94$), but $\tau = 1$~s puts the first zero
of \cref{eq:boxcar} at $1/\tau = 1$~Hz, the band centre, and collapses the
tracker to $0.32$ even at full sample rate. Third, on the delivery path, no
deep transfer-function notch at the cadence: a notch of $50\%$ depth, or one at
$0.6$~Hz near the low edge of the cadence band, costs at most $0.02$, but a notch of $90$--$100\%$ depth at
$1.0$--$1.2$~Hz roughly halves tracker power (to $0.45$--$0.63$) without erasing
it (\cref{app:meter}).

Against an honest operator, then, detection needs only a meter that samples at
$f_s \geq 2$~Hz (without an anti-alias filter), integrates over $\tau \leq 0.5$~s, and has no deep notch at the
cadence.

\section{Detection against a concealing operator}
\label{sec:frontier}
\Cref{sec:results} assumed a benign operator, who leaves the training cadence for the tracker to find. A concealing
operator can instead de-periodicise its schedule and smear the cadence away
(\cref{subsec:adversary}). In principle this is always possible. The concealing operator can insert idle time at random between iterations, or vary the work done in each, to remove any frequency for the Viterbi detector to identify. The question is the cost in training throughput the operator must incur in order to de-periodicise enough to hide from the Viterbi detector, and in particular whether any de-periodicisation strategy is both cheap and effective.

In this section we sweep all eight schedule strategies of \cref{tab:attacks}
and measure how much each one hides from the detectors. The
operator applies each strategy to its dominant training run, $P^{\mathrm{main}}$
in the aggregate of \cref{eq:aggregate}. The background of small trainings and
fine-tunings, $P^{\mathrm{tr}}_j$ and $P^{\mathrm{ft}}_k$, stays honest, and every cell is scored against the aggregate null of
\cref{eq:aggregate_null}. For each strategy we step its budget knob over the
range in \cref{tab:attacks}. We measure the detection rate of the Viterbi tracker and of the
fixed tests of \cref{subsec:comparators}, taking the better of the two fixed
tests at each cell. All rates are at false-alarm rate $10^{-2}$ with $n=200$
traces per class.

We measure the throughput cost incurred by four of the concealing strategies (period jitter, cadence drift, work variation and shape filling) on a single A100. We write the throughput cost as
\begin{equation}
	c \;=\; \frac{r_{\mathrm{honest}}}{r} - 1,
	\label{eq:throughput_cost}
\end{equation}
i.e. the fractional increase in wall-clock time per trained token, where $r$ is the
training throughput in tokens per second under the concealing strategy and
$r_{\mathrm{honest}}$ that of the same run without it.
For the strategies that only insert idle time, every iteration trains the same
number of tokens, so $c$ is the idle time added per iteration as a fraction of
the busy time. Work variation changes the number of tokens per iteration, so
there we measure $r$ directly in tokens per second. \Cref{app:sweep} gives
the measurement details.
We did not run the other four concealing strategies (phase slip,
relocation, harmonic smoothing and dilution) on hardware, so they have no
measured cost. 

\begin{figure*}[t]
	\centering
	\includegraphics[width=\linewidth]{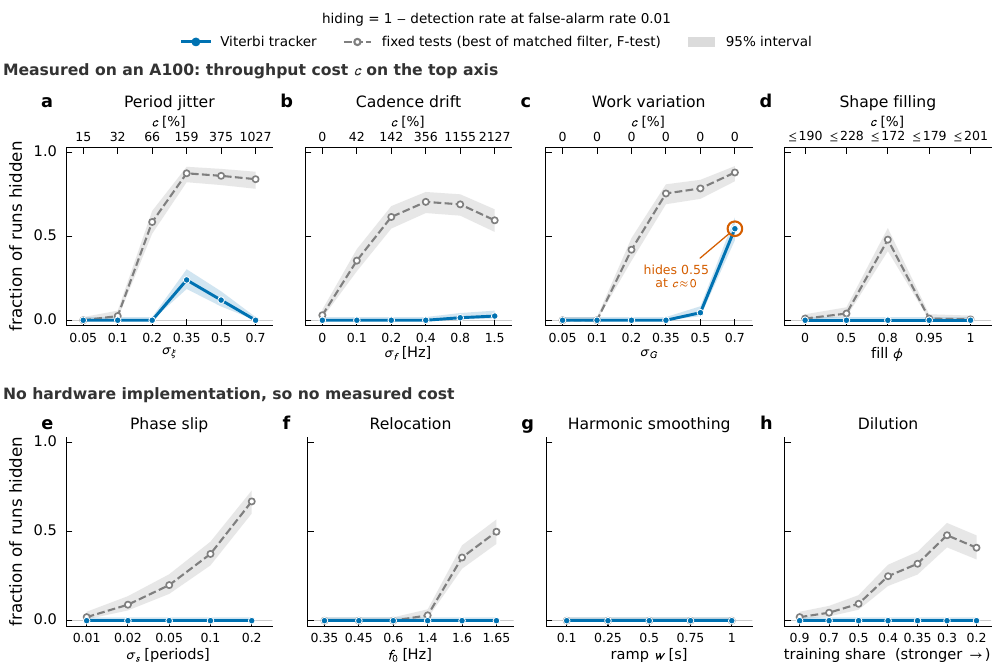}
	\caption{How much each schedule strategy hides, and what it costs. Each panel
	steps one strategy of \cref{tab:attacks} through its budget knob (bottom
	axis; settings evenly spaced, weakest on the left) and plots hiding, $1-$
	detection rate at false-alarm rate $10^{-2}$, for the Viterbi tracker (blue,
	solid) and the fixed tests (grey, dashed; the better of the matched filter
	and multitaper F-test per cell). Bands are Wilson $95\%$ intervals for
	$n=200$ traces per class. \textbf{a--d}, the four strategies run on a
	single A100. The top axis gives the measured throughput cost $c$ of each
	setting (\cref{eq:throughput_cost}); costs below $1\%$ in magnitude, zero
	within noise, are shown as $0$, and drift-free scheduling costs zero by
	definition. Shape-filling costs were measured on a jittered base and are
	upper bounds ($\leq$; \cref{app:sweep}).
	\textbf{e--h}, the four strategies we did not run on hardware, which therefore
	have no measured cost. The tracker detects almost every cell. The exception, ringed in
	\textbf{c}, is work variation at $\sigma_G = 0.7$, which hides $0.55$ from it
	at $c \approx 0$; idle-insertion jitter hides at most $0.24$, for
	$c = 159\%$. The meter strategy degrades the channel rather than the
	schedule and is treated in \cref{subsec:meter_spec}.}
	\label{fig:attack_curves}
\end{figure*}

\Cref{fig:attack_curves} plots hiding ($1-$ detection rate) against the
budget knob (and, where applicable, the measured throughput cost) of every concealing strategy. Against the fixed baseline detection methods, seven of the eight strategies hide a large fraction of
runs at their larger budgets; only harmonic smoothing hides nothing. Hiding
mostly grows with the budget, up to $0.88$ under both jitter and work
variation. Only work variation hides for free. It hides $0.88$ at $c \approx 0$,
while jitter and drift need $c = 159\%$ and $356\%$ to reach $0.88$
and $0.71$. Against the Viterbi tracker almost nothing hides. The tracker detects at least
$95\%$ of runs in $43$ of the $46$ cells. It detects every run in every cell of
phase slip, relocation, harmonic smoothing, shape filling and dilution, and at
least $97\%$ under drift. The three exceptions are two jitter settings and the
largest work variation. Jitter at $\sigma_\xi = 0.35$ hides $0.24$
for $c = 159\%$, and at $0.5$ hides $0.12$ for $c = 375\%$. Work variation at $\sigma_G = 0.7$ hides $0.55$ at $c \approx 0$,
though at every smaller $\sigma_G$ it hides at most $0.05$.

Independent measurements on real hardware, made with a board-level side
channel rather than a facility meter, show the same pattern of costs.
\citet{gargiulo2026workload} measure four evasion strategies on a single H200
--- chunked optimizer updates, dilution with decoy inference, throttling, and
diluted LoRA --- and find that they cost between $28$ and $69\%$ of
throughput. Each strategy inserts or substitutes
work, which is what makes it expensive, and none varies the real computation per
iteration, the one strategy we find cheap.

While the Viterbi detector is robust to the majority of concealing strategies, work variation at $\sigma_G = 0.7$ is the main exception; the operator can hide 55\% of $300$\,s observation windows at zero throughput cost. However, the strategy does have a cost in terms of training efficiency. The work variation strategy involves a few iterations accumulating many micro-steps, where about one in ten accumulates 64 micro-steps or more (\cref{eq:work_laws}). The long blocks raise the mean micro-step count from $8$ to $25$, so the operator takes about a third as many optimiser steps per token as honest training, and each step
on a larger batch does less per token. \citet{mccandlish2018empirical} model an
optimiser step on a batch of $B$ samples, with the learning rate tuned for that
batch, as making a fraction
\begin{equation}
	h(B) \;=\; \frac{B}{B + B_{\mathrm{noise}}}
	\label{eq:step_efficiency}
\end{equation}
of the progress of a full-gradient step, where the gradient noise scale
$B_{\mathrm{noise}}$ sets the critical batch size. For micro-step counts $G_i$
drawn independently, progress per sample is proportional to
$\mathbb{E}[h(G)]/\mathbb{E}[G]$, so the attack needs a fraction
\begin{equation}
	\Delta \;=\; \frac{h(G_0)/G_0}{\mathbb{E}[h(G)]/\mathbb{E}[G]} \;-\; 1
	\label{eq:token_penalty}
\end{equation}
more samples than honest training to reach the same loss, with $B$ and
$B_{\mathrm{noise}}$ counted in micro-batches; at $\sigma_G = 0.7$ it gives
$\Delta = 53$--$184\%$ for $B_{\mathrm{noise}}$ between $8$ and $128$.

We also measured the training efficiency penalty for the work variation strategy directly. We trained a $30$M-parameter GPT-style
language model on $400$M tokens of FineWeb-Edu \citep{penedo2024fineweb} on a
single A100, three seeds per schedule. The strategy uses the same micro-batches in
the same order as honest training ($G_0 = 8$), changing only how many are
accumulated into each step; the gradient noise scale we measured over these runs
is about $14$ micro-batches. We define the learning cost as the extra tokens
the strategy needs, i.e. the $400$M tokens the strategy trained on, divided by
the tokens honest training needs to reach the same final loss, minus one. At $\sigma_G = 0.7$ it is
$267 \pm 33\%$, cf.\ the predicted $163\%$ at the measured noise scale from \cref{eq:token_penalty}. We also tried a second form of work variation: draw $G_i$ lognormally
with spread $s$, with the mean held at the honest $8$. It takes as many
optimiser steps per token as honest training and never idles, so it should cost
no throughput. At $s = 1.0$ it hides $0.73$ from the tracker, more than the
$0.55$ of work variation at $\sigma_G = 0.7$, but costs $93 \pm 6\%$ in
training efficiency. Retuning the optimiser does not remove this cost. With
the learning rate, gradient clipping and $\beta_2$ tuned separately for honest
training and for the attacker, the lognormal form still costs $81 \pm 5\%$
(\cref{app:generator}).

\section{Discussion}
\label{sec:discussion}

\Cref{sec:results,sec:frontier} show that a Viterbi tracker detects synchronous training against a declared
inference workload, given a meter that samples at ${\geq}2$~Hz with ${\leq}0.5$~s
integration. The method is robust against the majority of concealing strategies, while the minority of strategies that hamper detection pay a cost in either throughput or learning efficiency. In this section we discuss the implications of these results for a verifier who is monitoring a facility for signs of training. This section is organised as follows. \Cref{subsec:runs} extends the short, per-window
detection rates of \cref{sec:results,sec:frontier} into the probability that a concealing operator is detected over a long training run. \Cref{subsec:identifiability} considers confounding, non-training workloads that imitate training and that the tracker also detects. \Cref{subsec:limitations} discusses limitations of the method and our analysis, and suggests avenues for future work.

\subsection{Detection probability for long runs}
\label{subsec:runs}

Every detection rate in \cref{sec:frontier} refers to a single $300$\,s window. However, a frontier training run lasts days to months. The verifier observes the power meter over these longer timescales. The quantity that both the verifier and the concealing operator are interested in is the probability of being detected at some point during these longer runs. In this section we compute this detection probability.  

For the majority of concealing strategies the Viterbi detector detects training in at least three quarters of windows, and in most settings in nearly all of them, so a run is easily caught. We therefore consider exclusively the work variation strategy, which was the only strategy that could hide a large fraction of windows from the tracker. We take two forms of work variation, the $\sigma_G = 0.7$ setting of \cref{tab:attacks} and the
lognormal variant introduced in \cref{sec:frontier}. We consider a training run to be a continuous $24$-hour trace of the aggregate of \cref{eq:aggregate}, cut into consecutive
$300$\,s windows and scored by the tracker. The run's cadence and background are
held fixed for the whole day, so the windows of one run are not independent
draws. A window counts as detected at false-alarm rate $10^{-2}$, as in
\cref{sec:frontier}. We use an anytime-valid test \citep{ramdas2023savi} to convert a sequence of Viterbi results on a sequence of windows into a decision about the run. Each window's score becomes a conformal p-value against $5000$ independent null windows, $p = (1 + \#\{\text{null scores} \ge V\})/5001$ \citep{vovk2005algorithmic}, each p-value
becomes an e-value $e = 1/(2\sqrt{p})$, and the test raises an alarm when the
running product of e-values first exceeds $1/\alpha$\footnote{By Ville's inequality
\citep{ville1939etude}, the false-alarm rate stays below $\alpha$ however long the
verifier watches and whenever it stops.}. We set $\alpha = 10^{-3}$. Across $2000$
null runs the empirical false-alarm rate after a day is $1.5 \times 10^{-3}$
($95\%$ interval $0.5$--$4.4 \times 10^{-3}$), consistent with $\alpha$. We also tested a simpler counting rule, which flags a run when the number of
windows detected by a fixed time would be improbable for a null facility. It also
keeps its false-alarm rate (two false alarms in $2000$ null runs after a day), but
the observation time must be fixed in advance, and it flags fewer runs.

\begin{figure*}[t]
	\centering
	\includegraphics[width=\linewidth]{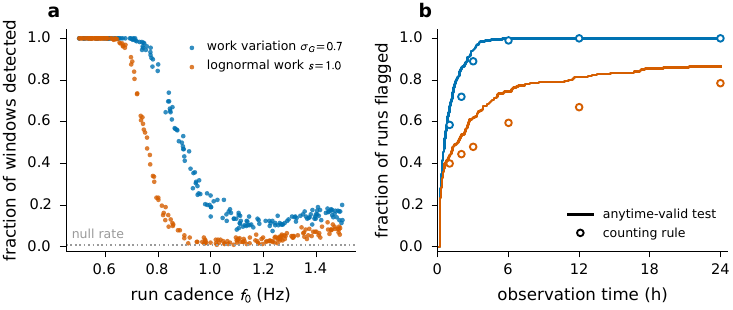}
	\caption{Detection of $200$ runs under each of two work-variation strategies,
	$\sigma_G = 0.7$ (blue) and lognormal (orange). Each run is one continuous
	$24$-hour trace of the aggregate, scored in $300$\,s windows.
	\textbf{a}, fraction of each run's windows detected at false-alarm rate
	$10^{-2}$, against the run's cadence. The dotted line is the null's $1\%$
	rate. \textbf{b}, fraction of runs flagged against observation time, at a
	run-level false-alarm rate of $10^{-3}$. Lines are the anytime-valid test,
	which may stop at any time; open circles are the counting rule, a separate
	fixed-time test at each marked time. Across $2000$ null runs the anytime-valid test raises
	$3$ false alarms in a day.}
	\label{fig:run_level}
\end{figure*}

Results are presented in \cref{fig:run_level}. Under the work variation strategy at $\sigma_G = 0.7$, runs with $f_0 \lesssim 0.8$\,Hz are detected in almost every window, and those above about $1.1$\,Hz in $8$--$20\%$. Even the most difficult of these runs is detected at several times the null's $1\%$ rate, so the test
catches it eventually given sufficient observation time (\cref{fig:run_level}b). It flags $68\%$ of runs within an hour, $95\%$ within three hours, and every run within six. For the lognormal work variation of \cref{sec:frontier}, runs with $f_0$ between about $0.9$ and $1.3$\,Hz are detected in only $1$--$7\%$ of windows, and $15$ of $200$ runs fall to $2\%$
or below, close to the null's rate. The test flags $44\%$ of these runs within an
hour, $75\%$ within six hours, and $87\%$ after a day. The remaining $13\%$
survive a day of observation. For $25$ of them the e-values do not grow on
average, so this test would not flag them however long it watched. A concealing operator can therefore hide a run, but only with
the lognormal strategy, only at some cadences, and at the $93\%$ learning cost
measured in \cref{sec:frontier}.

\subsection{Confounds}
\label{subsec:identifiability}

A detection by the Viterbi method establishes that the power trace carries a persistent, training-shaped
periodic schedule. However it does not establish that the computation is actually training. Ordinary operations could lead to similar periodicities that trip the detector, e.g. a non-ML kernel loop with a training-shaped cycle, a periodic
inference request generator, or a controller limit cycle, see \cref{app:rung2}. Instrumental lines that arise from the hardware itself are also possible. For example, on measured single-A100 traces the spectrum
contains a $0.31$--$0.45$\,Hz component due to the card's own power-management loop, which appears whenever the load varies in time but not at idle or under constant load.\footnote{This line need not grow in step with training at a facility meter. The
controllers of $N$ independent cards should add incoherently, growing roughly as
$\sqrt{N}$, while a synchronous training line grows as $N$, so the controller
line should fall relative to training at cluster scale.} Instrumental lines do not help a concealing operator hide training, but they could lead to otherwise compliant operators running inference being erroneously flagged. Compliant inference can also be periodic. A pool of servers dedicated to prefill under sustained load \citep{patel2024splitwise}, or offline batch inference such as bulk evaluation, repeats synchronous forward passes and transfers and may carry a training-like cycle.
Gravitational-wave searches face the same problem, and discard candidates that
coincide with a catalogue of known instrumental lines \citep{covas2018lines}. A
verifier could keep a similar list for each facility, with two differences.
Building it needs access to characterise the facility's hardware, and an
operator who knows the list can move its cadence into a listed band.

\subsection{Limitations}
\label{subsec:limitations}

The primary limitation of this work is that it uses synthetic data, generated
from a model of datacentre power draw (\cref{sec:scenario}). The training part
of that model (\cref{subsec:ko}) is nevertheless anchored to published
measurements. \citet{ko2025widearea} take
its parameters from published characterisations of datacentre and GPU power: an
oscillation near $1$\,Hz, compute phases that fill $55$--$80\%$ of each iteration,
and a ${\sim}30\%$ drop in draw during synchronisation. They report that the
profiles it generates agree qualitatively with public power traces and with the
experience of datacentre engineers. They could not test it quantitatively, because
facility-scale measurements are proprietary. The inference part of the model (\cref{subsec:null}) has weaker
support. \citet{ko2025widearea} give no inference model, so the one in
\cref{subsec:null} is our own. It follows the published behaviour of serving
systems, but we have not checked it against measured traces. The natural extension of this work is to replace the synthetic model with empirical multi-GPU training observed by an external meter. That study is planned but requires hardware and an experimental setup that are not currently available. Measurements from a single A100, read through NVML, are possible. However they are a poor stand-in for what a facility meter sees, for two reasons. First, one card has no communication phase. On a
single saturated A100 the iteration cadence is present but shallow: it modulates
a ${\sim}370$\,W draw by only a few watts (${\sim}0.4\%$ of board power); the deep
compute--communication contrast that makes the cadence legible comes from
synchronisation across many accelerators, so a single card is the case in which
it is weakest. Second, NVML reports one card's power from its own sensors, while
a facility meter sees the aggregate of many cards through the power-delivery
path. For the quantity a facility meter sees, the scenario model of
\citet{ko2025widearea} is therefore the better guide. We use the A100
measurements only where a single card is a fair proxy, e.g. the throughput cost of
the concealing strategies (\cref{sec:frontier}).

Several further limitations apply. (i) The observation map of \cref{eq:meter} is a
linear time-invariant first approximation. (ii) The tracker's threshold is set from simulated traces of our modelled null, so its false-alarm rate is guaranteed only if a real facility's null resembles that
model. A deployed verifier would instead calibrate on the facility's own
training-free traces or on per-trace surrogates (\cref{subsec:threshold}). (iii) Our estimate of the learning efficiency cost in \cref{sec:frontier} was only measured at a small scale with a $30$M-parameter model, one dataset, and
a limited search over concealing schedules. Whether
it holds at frontier scale, and against strategies we did not test, is an open question. (iv) The run-level results of \cref{subsec:runs} cover a single day of observation, and the
anytime-valid guarantee there assumes the window p-values are independent under
the null. The null runs support this for our modelled facility, but a real
facility's false alarms could be more correlated, for example through a
persistent background over weeks, so a deployed verifier would calibrate
against that facility's own windows. (v) We target synchronous, iteration-structured training whose cadence lies in the searched band; a run that
moves its cadence outside the band or abandons synchronous iteration is outside
the population we test. (vi) The declared workload need not be pure inference; a facility may be permitted fine-tuning or training below a threshold, whose lines our detector would also find, and telling which run a line belongs to and how large it is needs the integrated-energy channel we exclude, which bounds the amount of computation only weakly against an adversary \citep{kimpson2026power}. (vii) Our claims on the robustness of the Viterbi method to concealing strategies and cost budgets are limited to those strategies we considered; there may be alternative strategies not considered here that the operator could employ. One important example  is energy storage between the accelerators and the meter. An energy store sized to the cadence swing could discharge during each compute phase and recharge during each communication phase, so that the meter sees a flat load. This costs the
operator no throughput, but they would incur additional costs such as purchasing power electronics sized for the full swing, maintenance of a store that wears from charging and
discharging once per iteration, and energy lost on every cycle.
Power smoothing of this kind is a grid mitigation for training load in its own
right \citep{ko2025widearea}, so it may be installed for reasons unrelated to
concealment. The verifier can respond by moving the meter, or adding a second one, between the
store and the accelerators, where the load is not yet smoothed. A meter on the
store itself also works: to flatten the load, the store must charge and discharge
at the training cadence, so its own power flow carries the line. Both meters stay
off-chip if the store sits at the facility or row level. If it sits inside the
racks, the meter must move closer to the chips than we assume.

\section{Conclusion}
\label{sec:conclusion}

We have shown that a passive, off-chip power meter can detect synchronous training hidden in a declared inference workload. A Viterbi detector finds the iteration cadence of training, and is robust to many different concealing strategies. Only varying the real work per iteration weakens the detector. Over a long training run, an anytime-valid test catches most attacked runs within a day. Hiding training is
expensive: the strongest attack we found evades detection in about one run in
seven after a day, but needs $93\%$ more compute to reach the same loss. 

\section*{Acknowledgements}

TK and EG thank the Machine Alignment, Transparency \& Security (MATS) programme---its
directors, organisers, funders, and staff---who made it possible for us to
work on this project and who provided invaluable resources and support
throughout.

\bibliography{references}
\bibliographystyle{icml2026}

\newpage
\appendix
\crefalias{section}{appendix}
\crefname{appendix}{Appendix}{Appendices}
\Crefname{appendix}{Appendix}{Appendices}
\onecolumn

\section{Optimality ceiling}
\label{app:ceiling}

We can compare the Viterbi tracker with an
optimality ceiling: an approximation to the Neyman--Pearson optimal test between
the training and null generators. The optimal test compares how likely a trace
is under $\mathcal{H}_1$ and under $\mathcal{H}_0$. Our generators can simulate
traces from either class but cannot give the probability of a given trace, so
we approximate. We summarise each class by its average power spectrum in the
search band, estimated from simulated traces, and ask which spectrum better
explains the periodogram of the trace (a Whittle likelihood ratio). The training
cadence is unknown, so the training likelihood is marginalised over a bank of $61$
templates $\bar S_m$, one per value of $f_0$ with a uniform prior, each estimated from $2000$ simulated traces
with drift levels drawn from the whole $\sigma_f$ axis. For the in-band
periodogram $I(f)$ of the full trace, the statistic is
\begin{equation}
	\Lambda \;=\; \log \frac{1}{61}\sum_{m=1}^{61} e^{\ell(\bar S_m)} \;-\; \ell(\bar S_0),
	\qquad
	\ell(S) \;=\; -\sum_{f \in \mathcal{B}} \Bigl[\ln S(f) + \frac{I(f)}{S(f)}\Bigr],
	\label{eq:whittle}
\end{equation}
with $\bar S_0$ the null spectrum. The ceiling therefore
knows no more about the drift than the tracker does. Each template fixes the
dominant run's cadence over a random background, and the null spectrum is
estimated from the null aggregate. The traces that build these spectra are drawn
from a stream disjoint from the traces we score, and the ceiling is applied to
exactly the populations of \cref{fig:drift_ceiling}.

The ceiling is a benchmark, not a usable detector. It knows the exact spectra of
both classes, which a verifier monitoring a real facility does not. The tracker
assumes only a slowly wandering line, and the ceiling measures how much of the
available detection power it recovers.

The ceiling detects every trace at every drift (\cref{fig:drift_ceiling}). The
training line and the null are therefore separable in principle from a
$300$~s trace, and the tracker's shortfall, at most $0.04$ at
$\sigma_f = 1.5$~Hz, is a limit of the detector rather than of the problem. One
way to close it is known. Under the hidden-Markov wander model the tracker
already assumes, the Viterbi score follows the single best path, while the
Neyman--Pearson statistic sums over all frequency paths with the forward
algorithm \citep{streit1990frequency}. We computed it, with the same contrast
map and wander penalty, normalising the penalty into transition probabilities,
on fresh populations of $300$\,s windows thresholded at false-alarm rate
$10^{-2}$. It matches the tracker on honest and drifting runs, and does
somewhat better under work variation: detection rises from $0.45$ to $0.61$ at
$\sigma_G = 0.7$ and from $0.28$ to $0.36$ under the lognormal form. Most of
the hiding of \cref{sec:frontier} therefore survives the statistic that is
optimal under the tracker's own model.

The ceiling is optimal only under the Whittle model, which uses the power
spectrum alone and so ignores the phase relations between harmonics and any
non-Gaussian structure. The true optimum can only be higher. Since the ceiling
already detects every trace, this changes nothing here.

The same comparison on a single training run
against inference alone, with no background load, gives tracker detection rates
of $1.00$ up to $\sigma_f = 0.8$~Hz and $0.98$ at $1.5$~Hz, within $0.02$ of
the aggregate at every drift. At the default shares the background neither helps nor hinders the
tracker.

\section{Fixed-frequency baselines}
\label{app:baselines}

This appendix defines the multitaper F-test of \cref{subsec:comparators}. Taper
the mean-removed trace with $K$ discrete prolate spheroidal sequences $v_k[n]$
of time--bandwidth product $NW$, let $y_k(f)$ be the Fourier transform of
$v_k[n]\,x[n]$, and let $U_k = \sum_n v_k[n]$ be the $k$-th taper's gain at
zero frequency. The least-squares amplitude of a line at $f$ and the harmonic F
statistic are \citep{thomson1982multitaper, percival1993spectral}:
\begin{equation}
	\hat\mu(f) \;=\; \frac{\sum_k U_k\, y_k(f)}{\sum_k U_k^2},
	\qquad
	F(f) \;=\; (K-1)\,
	\frac{\lvert\hat\mu(f)\rvert^2 \sum_k U_k^2}
	{\sum_k \bigl\lvert y_k(f) - \hat\mu(f)\,U_k \bigr\rvert^2},
	\label{eq:thomson_f}
\end{equation}
the power in the fitted line over the power the tapers see around it. We use
$K = 7$ and $NW = 4$, on a frequency grid zero-padded to four times the next
power of two above the trace length.

At a single fixed frequency, and for a stationary null whose spectrum is smooth
near $f$, $F(f)$ follows an $F(2, 2K-2)$ distribution. The test statistic of
\cref{eq:ftest} is instead the maximum of $F$ over the search band, so its null
distribution needs a correction for the number of frequencies searched. The
analytic correction does not work. A Bonferroni correction over the number of
independent frequency cells sets thresholds that are too low: the realised
false-alarm rate exceeds the nominal one by a factor of about $3.6$--$6$ on every
null we tried, because $F$ fluctuates on scales finer than a frequency cell.
Real power traces would also break the smooth-spectrum assumption. We therefore
threshold both fixed tests empirically, as for the tracker
(\cref{subsec:threshold}).

\section{Generator and detector details}
\label{app:generator}

This appendix gives additional details on the workload generator, the power
meter, the adversarial strategies, the detectors, and the learning-efficiency
runs. The code in the repository is the reference implementation.

\paragraph{Workload generator.} Workloads are synthesised on a $20$\,Hz grid.
Each sample takes the level of the phase it falls in (\cref{eq:ko}), plus
independent Gaussian intra-phase noise, so phase boundaries are quantised to
the grid. Workload power is clipped at zero, so it never falls below the static
overhead. In the aggregate of \cref{eq:aggregate}, the dominant run is scaled
by $w_{\mathrm{main}}/(w_{\mathrm{main}} + w_{\mathrm{tr}} + w_{\mathrm{ft}})$,
and each of the $N_{\mathrm{tr}}$ small trainings by $w_{\mathrm{tr}}/
[N_{\mathrm{tr}}(w_{\mathrm{main}} + w_{\mathrm{tr}} + w_{\mathrm{ft}})]$, and
likewise for fine-tuning. In the inference null of \cref{eq:null}, $b(t)$ is
white Gaussian noise with every Fourier coefficient outside
$[f_b^{\mathrm{lo}}, f_b^{\mathrm{hi}}]$ set to zero, rescaled to unit RMS.
$u(t)$ is a discrete AR(1) process with lag-one coefficient
$e^{-\Delta t/\tau_u}$, standardised to zero mean and unit variance. Each
inference sample also carries independent Gaussian noise of standard deviation
$0.035\,\kappa_\nu$, the analogue of training's intra-phase noise.

\paragraph{Meter.} The low-pass is a second-order Butterworth filter, and the
notch a second-order notch of quality factor $Q$ whose depth $d$ blends the
signal toward its notched version, $x \to x + d\,(\mathrm{notch}(x) - x)$. Both
are applied forward and backward, so they have zero phase. After the boxcar,
each meter reading at $t_n = n/f_s$ is the most recent sample of the $20$\,Hz
grid, with no anti-alias filter. AR(1) meter noise has marginal standard deviation
$\sigma_\eta$ and lag-one coefficient $e^{-1/(f_s \tau_\eta)}$. The four hostile
channels of the meter strategy (\cref{tab:attacks}) are:
\begin{itemize}
	\item a full-depth notch at $1.0$\,Hz with $Q = 1.5$;
	\item a $1$\,s boxcar sampled at $1$\,Hz;
	\item a controller limit cycle of $30$\,W at $0.38$\,Hz, $40\%$ duty and
	$0.05$\,Hz wander;
	\item AR(1) noise of $\sigma_\eta = 12$\,W with $\tau_\eta = 2$\,s, plus a
	$20$\,W baseline wander with a $60$\,s correlation time.
\end{itemize}
The other three channels carry white noise of $\sigma_\eta = 4$\,W. The
effective signal-to-noise ratio of \cref{fig:scenario_meter} is
$\hat S_{\mathrm{sig}}(f_0)/\langle \hat S_{\mathrm{flat}} \rangle_{\mathcal{B}}$,
normalised to the nominal channel. The numerator is the multitaper spectrum of
the metered training trace at the known $f_0$. The denominator is the
spectrum averaged over the band (up to $f_s/2$) when the same meter, with the
same noise realisation, observes a constant load.

\paragraph{Concealing strategies.} Work variation splits the honest up-phase
into $G_0$ micro-steps of length $t_\mu = \rho_i T_i/G_0$. Each iteration
draws a real-valued count and rounds it stochastically to an integer
$G_i \geq 1$:
\begin{equation}
	\tilde G_i = \frac{G_0}{\max(1 + \zeta_i,\, 0.05)},\quad \zeta_i \sim \mathcal{N}(0, \sigma_G^2),
	\qquad\text{or}\qquad
	\tilde G_i = \min\!\bigl(G_0\, e^{s z_i - s^2/2},\, G_0/0.05\bigr),\quad z_i \sim \mathcal{N}(0,1),
	\label{eq:work_laws}
\end{equation}
for the $\sigma_G$ and lognormal forms. The up-phase lasts $G_i t_\mu$, and the
down-phase keeps its honest duration. The reciprocal form has a heavy right
tail. At $\sigma_G = 0.7$, $G_i \geq 64$ requires $\zeta_i \leq -0.875$, which
happens in $11\%$ of iterations; this tail raises the mean count from $8$ to $25$.
The lognormal form keeps the mean at $G_0$. The other strategies change the
generator as follows:
\begin{itemize}
	\item Phase slip moves the clock after each iteration by
	$\mathcal{N}(0, \sigma_s^2 T_i^2)$, without changing the target period.
	\item Shape filling raises each down-phase level by $\phi$ times its gap
	to that iteration's up-phase level.
	\item Harmonic smoothing convolves the level waveform with a Hann kernel of
	total width $w$ before the intra-phase noise is added.
	\item Dilution sets the dominant run's share of peak power and keeps the
	background's $w_{\mathrm{tr}} : w_{\mathrm{ft}}$ ratio, so the freed share
	goes to the background trainings and fine-tunings.
	\item Relocation pins the dominant run's $f_0$.
\end{itemize}

\paragraph{Detectors.} The tracker's frames use a Tukey window with taper
fraction $0.25$. A frame lasts $\max(64, 16 f_s)$ samples, which is $16$\,s at
$f_s \geq 4$\,Hz but $32$\,s at $2$\,Hz and longer below. Only bins at or below
$f_s/2$ enter, and $V = 0$ if no bin of $\mathcal{B}$ remains, as at
$f_s = 0.5$\,Hz. The matched filter's Welch segments are
$\min(N, 1024)$ samples long, so at $f_s \leq 2$\,Hz it is a single periodogram
of the whole trace. Its median runs over all frequencies from $0$ to $f_s/2$.

\paragraph{Thresholds.} Each window-level detection rate at false-alarm rate
$10^{-2}$ is read from the ROC of the $n = 200$ scored traces of each class. It
is the largest detection rate whose threshold lets at most $2$ of the $200$ null
traces through. The Wilson intervals of \cref{fig:attack_curves} are for the
detection rate at that threshold and do not include the uncertainty in the
threshold itself.

\paragraph{Learning-efficiency runs.} The $30$M-parameter model trains with
AdamW (peak learning rate $10^{-3}$, $\beta = (0.9, 0.95)$, weight decay
$0.1$, gradient clipping at $1.0$), and its learning-rate schedule is indexed
by tokens, so every schedule of $G_i$ follows the same schedule per token. The
lognormal attacker was also run under six other recipes: peak learning rate
$5\times 10^{-4}$ or $2\times 10^{-3}$, clipping off, $\beta_2 = 0.99$, and
combinations of these. The same changes were applied to honest training. Against
the best honest recipe, the best attacker recipe costs $81 \pm 5\%$ (three
seeds). Scaling each step's learning rate by $\sqrt{G_i/G_0}$ instead raises the
cost to $116 \pm 5\%$. The $\sigma_G = 0.7$ attacker was not retuned.

\section{Meter specification details}
\label{app:meter}

This appendix gives two results behind the minimum meter specification of
\cref{subsec:meter_spec}: the notch sweep, and a check that the specification
does not depend on the background load.

\begin{figure}[h!]
	\centering
	\includegraphics[width=0.6\columnwidth]{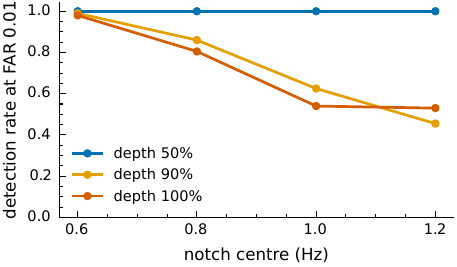}
	\caption{Notch sweep: Viterbi detection rate at empirical false-alarm
	rate $10^{-2}$ versus notch centre, one line per notch depth, at $20$~Hz
	sampling with no integration window, the honest aggregate against the null
	aggregate ($n=200$ per class). A deep notch near the band centre roughly
	halves detection; none removes it.}
	\label{fig:meter_boundary_notch}
\end{figure}

\Cref{fig:meter_boundary_notch} shows the tracker's
detection rate as a delivery-path notch moves across the training cadence band
of \cref{tab:params} ($0.5$--$1.5$~Hz), with the meter otherwise nominal. The
notch has quality factor $Q = 1.5$ (\cref{tab:meter}), so it is broad: about
$0.7$~Hz wide at $1$~Hz. A notch of $50\%$ depth costs nothing at any centre,
and one at $0.6$~Hz, near the low edge of the band, costs at most $0.02$. Deeper
notches nearer the band centre cost more. At $0.8$~Hz the tracker falls to
$0.86$ ($90\%$ depth) and $0.80$ ($100\%$), and at $1.0$--$1.2$~Hz to
$0.45$--$0.63$. No notch in the sweep removes detection entirely.

The same sweep on a single training run against
inference alone, with no background load, changes the tracker's detection rate
by at most $0.09$ in any cell and leaves the region where detection holds
($f_s \ge 2$~Hz and $\tau \le 0.5$~s) unchanged. As for the drift sweep
(\cref{app:ceiling}), the background at the default shares does not move the
result.

\section{Throughput-cost measurements}
\label{app:sweep}

This appendix records how the sweep of \cref{sec:frontier} was run and how the
throughput costs of \cref{fig:attack_curves} were measured. The sweep applies
each strategy of \cref{tab:attacks} to the dominant training run of the
aggregate and scores it against the null aggregate, with $n=200$ traces per
class over the $46$ schedule cells of \cref{fig:attack_curves} (plus the four
meter channels of \cref{tab:attacks}), every rate read at false-alarm rate
$10^{-2}$.

The throughput cost $c$ of \cref{eq:throughput_cost} comes from a campaign on a
single A100. We use only its
throughput measurements. We do not use the size of the power line it measured,
since a single card is a poor guide to what a facility meter sees
(\cref{subsec:limitations}), so the costs do not depend on that caveat. The
card trains an untrained decoder-only transformer on random tokens, with a
fixed number of gradient-accumulation micro-steps per step, so it runs at a
busy-loop cadence set by its compute. Each strategy is realised on this loop.
Jitter and drift insert sleep pads after each step. A pad can only lengthen a
period, so the target cadence is derated below the busy-loop cadence by three
standard deviations of the commanded excursion, and periods drawn faster than
that are clipped at zero pad. The derating is part of the measured cost. Work
variation instead draws each step's micro-step count, as in
\cref{app:generator}, and never sleeps. Shape filling replaces the sleep in
each pad with matrix-multiply bursts, chopped at a randomised $15$--$35$\,ms
quantum and at the duty cycle that raises the pad's power by the fraction
$\phi$ toward the training draw. It was measured on a $\sigma_\xi = 0.35$
jitter base, so its costs include that jitter's padding and are upper bounds. Each
cost is the mean over eight A100 traces at that setting. Work variation and
jitter at level $0.7$, and drift at $0.1$, $0.4$, $0.8$ and $1.5$~Hz, were
measured in a later run on the same hardware. The drift costs are
heavy-tailed: at $0.8$ and $1.5$~Hz the mean exceeds the median by factors of
$1.5$ and $2$. For work variation the campaign records the throughput deficit
$d = 1 - r/r_{\mathrm{honest}}$ rather than $c$. The two are related by
$c = d/(1-d)$ and differ by less than $0.01$ percentage points at the measured
$|d| < 1\%$, so we report $d$ directly.

\section{Non-training loads that score as training}
\label{app:rung2}

\Cref{subsec:identifiability} notes that a load can carry a training-shaped
power schedule without training. This appendix tests that with loads whose
training status we know, using $n=200$ traces of $300$~s per class.

We go beyond the single Viterbi score and give the verifier a
classifier. The populations are single workloads without the background
(a training run of \cref{eq:ko} against the inference null of \cref{eq:null}).
Each trace is reduced to eight features computed from the trace
alone, each oriented so that larger is more training-like:
\begin{itemize}
	\item the multitaper F-test of \cref{eq:ftest};
	\item a harmonic comb, which combines the F-test p-values at
	$f_0, 2f_0, \dots$ by Fisher's method and maximises over $f_0$;
	\item the spectral matched filter of \cref{eq:matched};
	\item the Viterbi score of \cref{eq:viterbi};
	\item two cyclostationary (Dandawat\'e--Giannakis) statistics, one at a
	fixed nominal cycle frequency and one after resampling the trace onto the
	phase of the Viterbi path;
	\item the smoothness of the Viterbi path from frame to frame;
	\item how well each recovered cycle correlates with the mean waveform when
	the trace is folded on the path's phase.
\end{itemize}
Three decision rules separate training from inference:
\begin{itemize}
	\item a fixed score that sums the features after standardising each on the
	null, with unit weights and the fixed-frequency cyclostationary feature
	left out;
	\item a logistic regression on all eight features, cross-validated over
	five folds;
	\item a random forest on waveform-shape features of the power trace, after
	\citet{rahman2026telemetry}.
\end{itemize}
All three separate the classes perfectly, with area under the ROC curve
$1.00$. They keep that score when the population is shifted: device headroom,
trace length, cadence band, an added controller line, or heavy coloured
meter noise.

We then fix each rule's threshold at a false-positive rate of $10^{-2}$ and
score seven control loads (\cref{tab:rung2_controls}). Five are not training:
\begin{itemize}
	\item a decoy that runs the full forward pass, backward pass and optimiser
	step and then discards the update, which is honest training in the
	generator;
	\item a computation of gradients alone, which is training with the
	optimiser step's swing removed ($\mu_\Delta = 0.05$, $\sigma_\Delta = 0.02$);
	\item a non-ML kernel loop with a training-shaped cycle, which is training
	with $0.5$\,s ramps on its transitions;
	\item a power-controller limit cycle, a $0.4$\,Hz square wave of $40\%$
	duty and amplitude $0.3$ on a base of $0.5$;
	\item a periodic inference request generator, which is the inference null
	multiplied by a $\pm 50\%$ square envelope at $0.9$\,Hz.
\end{itemize}
Every rule scores all five as training, in at least $97\%$ of traces. Two
loads are training. A run diluted to half the meter's load by unrelated
co-resident work is caught by every rule. An asynchronous run whose cadence
drifts ($\sigma_f = 1.2$\,Hz) and slips ($\sigma_s = 0.15$ periods) is caught
in $47$--$100\%$ of traces.

\begin{table}[h!]
	\centering
	\small
	\caption{Fraction of each control load scored as training at a
	false-positive rate of $10^{-2}$, for each decision rule. All five non-training
	loads score as training.}
	\label{tab:rung2_controls}
	\begin{tabular}{@{}lcccc@{}}
		\toprule
		Control load & Training? & Linear & Logistic & Forest \\
		\midrule
		Discarded-update decoy & no & 1.00 & 1.00 & 1.00 \\
		Gradient-only & no & 0.97 & 1.00 & 1.00 \\
		Non-ML kernel loop & no & 1.00 & 1.00 & 1.00 \\
		Controller cycle & no & 1.00 & 1.00 & 1.00 \\
		Periodic inference & no & 1.00 & 1.00 & 1.00 \\
		Co-resident mixture & yes & 1.00 & 1.00 & 1.00 \\
		Asynchronous training & yes & 0.59 & 0.47 & 1.00 \\
		\bottomrule
	\end{tabular}
\end{table}

The meter sees only the power
schedule. A verifier's decision is some function of the trace, so two loads that
produce traces with the same distribution are flagged with the same
probability, by any detector. The decoy reproduces the alternation of compute
and communication that makes up the training schedule, so it produces the same
trace as training without training a model. Amplitude does not help either: the
power drawn is the rate of operations times the energy per operation, and the
meter sees only the product. A higher sampling rate does not help. A classifier
reading a $10$~MHz board-level probe separates training from inference by the
power signature of the optimiser update \citep{gargiulo2026workload}, but the
decoy runs that update and only discards its result. A detection therefore
shows that a training-shaped schedule is running, not that a model is being
trained. For the verifier the practical concern is the benign loads in
\cref{tab:rung2_controls}, such as the controller cycle and the kernel loop,
which could flag a compliant facility; a concealing operator has no reason to
run a decoy (\cref{subsec:adversary}).

\end{document}

%% file: fig_power_chain_v2.tex
\begin{tikzpicture}[
	font=\sffamily\scriptsize,
	card/.style={draw=#1, fill=white, rounded corners=1.5pt, line width=0.55pt,
		align=center, inner sep=2.5pt, minimum height=1.0cm},
	sum/.style={draw=black!75, fill=white, circle, line width=0.55pt,
		inner sep=0pt, minimum size=3.6mm},
	sig/.style={-{Stealth[length=1.7mm, width=1.3mm]}, line width=0.6pt, black!75},
	dsc/.style={text=black!60},
	badge/.style={circle, fill=okverm, text=white, font=\sffamily\bfseries\tiny,
		inner sep=0pt, minimum size=3.0mm},
	hdr/.style={font=\sffamily\bfseries\tiny, anchor=west},
	tap/.style={densely dotted, black!45, line width=0.45pt},
	ax/.style={black!35, line width=0.4pt},
	sk/.style={okblue, line width=0.7pt, line join=round},
	scap/.style={font=\sffamily\scriptsize, text=black!75, anchor=north, align=center},
]
\fill[okorange!9, rounded corners=3pt] (0, -1.35) rectangle (7.45, 2.4);
\fill[okblue!8,   rounded corners=3pt] (7.6, -1.35) rectangle (17.1, 2.4);
\node[hdr, text=okorange!60!black] at (0.12, 2.17) {OPERATOR CONTROLS};
\node[hdr, text=okblue!70!black]   at (7.72, 2.17) {VERIFIER CONTROLS};

\node[card=okorange, text width=1.95cm] (dom) at (1.17, 0.6)
	{\textbf{Dominant run}\\ \textcolor{black!60}{training ($\mathcal{H}_1$) or}\\
	 \textcolor{black!60}{inference ($\mathcal{H}_0$)}};
\node[card=okorange, text width=1.95cm] (bg) at (1.17, -0.62)
	{\textbf{Background}\\ \textcolor{black!60}{small trainings,}\\
	 \textcolor{black!60}{fine-tunings}};
\node[card=okorange, minimum height=0.5cm] (p0) at (2.85, 1.5)
	{\textbf{Static overhead} $P_0$};
\node[sum] (s1) at (2.62, 0) {};
\node[card=okorange, text width=2.3cm] (hdel) at (5.1, 0)
	{\textbf{Delivery path} $h_{\mathrm{del}}$\\ \textcolor{black!60}{PSU, PDU, storage}\\
	 \textcolor{black!60}{low-pass, notch}};
\node[card=okblue, text width=2.1cm] (hins) at (9.1, 0)
	{\textbf{Meter} $h_{\mathrm{ins}}$, $f_s$\\ \textcolor{black!60}{integrate, calibrate,}\\
	 \textcolor{black!60}{sample}};
\node[sum] (s2) at (12.1, 0) {};
\node[card=okblue, minimum height=0.5cm, text width=3.4cm] (eta) at (12.1, 1.5)
	{\textbf{Additive terms} $\eta$\\
	 \textcolor{black!60}{meter noise, baseline wander,}\\
	 \textcolor{black!60}{controller line}};
\node[card=okblue, text width=1.85cm] (ver) at (16.0, 0)
	{\textbf{Verifier}\\ \textcolor{black!60}{detector}\\ \textcolor{black!60}{(\cref{sec:methods})}};
\foreach \s in {s1, s2} {
	\draw[black!75, line width=0.55pt] ([xshift=-1mm]\s.center) -- ([xshift=1mm]\s.center);
	\draw[black!75, line width=0.55pt] ([yshift=-1mm]\s.center) -- ([yshift=1mm]\s.center);
}

\draw[sig] (dom.east) -- (s1);
\draw[sig] (bg.east) -- (s1);
\draw[sig] (p0.south -| s1) -- (s1);
\draw[sig] (s1) -- node[above, pos=0.38] {$P_{\mathrm{fac}}$} (hdel);
\draw[sig] (hdel) -- (hins);
\draw[sig] (hins) -- (s2);
\draw[sig] (eta) -- (s2);
\draw[sig] (s2) -- node[above, pos=0.73] {$P_{\mathrm{meter}}$} (ver);

\node[badge] at (dom.north east)  {S};
\node[badge] at (hdel.north east) {M};
\node[badge] at (hins.north east) {M};
\node[badge] at (eta.north east)  {M};

\foreach \x in {3.3, 7.05, 10.8, 14.55} {
	\fill[black!75] (\x, 0) circle (0.75pt);
	\draw[tap] (\x, 0) -- (\x, -1.62);
	\draw[ax] (\x-1.45, -1.72) -- (\x-1.45, -2.95) -- (\x+1.45, -2.95);
}
\begin{scope}[shift={(1.85, -2.95)}]
	\draw[sk] (0.1, 0.95) \foreach \k in {0,1,2,3} {
		-- ({0.1 + \k*0.675 + 0.405}, 0.95) -- ({0.1 + \k*0.675 + 0.405}, 0.3)
		-- ({0.1 + (\k+1)*0.675}, 0.3) -- ({0.1 + (\k+1)*0.675}, 0.95) };
\end{scope}
\begin{scope}[shift={(5.6, -2.95)}]
	\draw[sk, domain=0.1:2.8, samples=90, smooth]
		plot (\x, {0.62 + 0.2*sin(360*(\x-0.1)/0.675 + 60)});
\end{scope}
\begin{scope}[shift={(9.35, -2.95)}]
	\draw[okblue!30, line width=0.6pt, domain=0.1:2.8, samples=90, smooth]
		plot (\x, {0.62 + 0.2*sin(360*(\x-0.1)/0.675 + 60)});
	\foreach \k in {0,...,15} {
		\pgfmathsetmacro{\xx}{0.1 + \k*0.17}
		\pgfmathsetmacro{\yy}{0.62 + 0.2*sin(360*(\xx-0.1)/0.675 + 60)}
		\draw[okblue!45, line width=0.35pt] (\xx, 0.08) -- (\xx, \yy);
		\fill[okblue] (\xx, \yy) circle (1.05pt);
	}
\end{scope}
\pgfmathsetseed{7}
\begin{scope}[shift={(13.1, -2.95)}]
	\draw[okblue!30, line width=0.6pt, domain=0.1:2.8, samples=90, smooth]
		plot (\x, {0.62 + 0.2*sin(360*(\x-0.1)/0.675 + 60)});
	\foreach \k in {0,...,15} {
		\pgfmathsetmacro{\xx}{0.1 + \k*0.17}
		\pgfmathsetmacro{\yy}{0.62 + 0.2*sin(360*(\xx-0.1)/0.675 + 60) + 0.09*rand}
		\coordinate (n\k) at (\xx, \yy);
	}
	\draw[okblue!70, line width=0.4pt] (n0) \foreach \k in {1,...,15} { -- (n\k) };
	\foreach \k in {0,...,15} { \fill[okblue] (n\k) circle (1.05pt); }
\end{scope}
\node[scap] at (3.3,   -3.05) {$P_{\mathrm{fac}}$: sharp training cadence};
\node[scap] at (7.05,  -3.05) {after delivery: smoothed};
\node[scap] at (10.8,  -3.05) {after meter: integrated, sampled};
\node[scap] at (14.55, -3.05) {$P_{\mathrm{meter}}$: with noise $\eta$};
\node[font=\sffamily\tiny, text=black!45, anchor=west] at (1.88, -2.82) {time $\rightarrow$};

\node[badge] at (5.55, -3.9) {S};
\node[anchor=west, text=black!75] at (5.7, -3.9) {schedule attacks, dilution};
\node[badge] at (9.15, -3.9) {M};
\node[anchor=west, text=black!75] at (9.3, -3.9) {meter strategy};
\node[anchor=east, text=black!75] at (5.35, -3.9) {Adversary acts at:};
\end{tikzpicture}

%% file: main.bbl
\begin{thebibliography}{31}
\providecommand{\natexlab}[1]{#1}
\providecommand{\url}[1]{\texttt{#1}}
\expandafter\ifx\csname urlstyle\endcsname\relax
  \providecommand{\doi}[1]{doi: #1}\else
  \providecommand{\doi}{doi: \begingroup \urlstyle{rm}\Url}\fi

\bibitem[Baker et~al.(2025)Baker, Kulp, Marks, Brundage, and
  Heim]{baker2025verifying}
Baker, M., Kulp, G., Marks, O., Brundage, M., and Heim, L.
\newblock Verifying international agreements on {AI}: Six layers of
  verification for rules on large-scale {AI} development and deployment, July
  2025.
\newblock URL \url{https://arxiv.org/abs/2507.15916}.

\bibitem[Bayley et~al.(2019)Bayley, Messenger, and Woan]{bayley2019soap}
Bayley, J., Messenger, C., and Woan, G.
\newblock Generalized application of the {Viterbi} algorithm to searches for
  continuous gravitational-wave signals.
\newblock \emph{Physical Review D}, 100:\penalty0 023006, 2019.
\newblock \doi{10.1103/PhysRevD.100.023006}.

\bibitem[Covas et~al.(2018)Covas, Effler, Goetz, Meyers, Neunzert,
  et~al.]{covas2018lines}
Covas, P.~B., Effler, A., Goetz, E., Meyers, P.~M., Neunzert, A., et~al.
\newblock Identification and mitigation of narrow spectral artifacts that
  degrade searches for persistent gravitational waves in the first two
  observing runs of {Advanced LIGO}.
\newblock \emph{Physical Review D}, 97:\penalty0 082002, 2018.
\newblock \doi{10.1103/PhysRevD.97.082002}.

\bibitem[Gargiulo \& Kulp(2026)Gargiulo and Kulp]{gargiulo2026workload}
Gargiulo, S. and Kulp, G.
\newblock Workload identification with physical side channels for {AI}
  governance, 2026.
\newblock URL \url{https://arxiv.org/abs/2609.00309}.

\bibitem[Grattafiori et~al.(2024)Grattafiori, Dubey, Jauhri, Pandey, Kadian,
  et~al.]{grattafiori2024llama3}
Grattafiori, A., Dubey, A., Jauhri, A., Pandey, A., Kadian, A., et~al.
\newblock The {Llama} 3 herd of models, 2024.
\newblock URL \url{https://arxiv.org/abs/2407.21783}.

\bibitem[Heim \& Koessler(2024)Heim and Koessler]{heim2024thresholds}
Heim, L. and Koessler, L.
\newblock Training compute thresholds: Features and functions in {AI}
  regulation, 2024.
\newblock URL \url{https://arxiv.org/abs/2405.10799}.

\bibitem[Jain et~al.(2024)Jain, Tran, Chen, Sinclair, and
  Venkataraman]{jain2024pal}
Jain, R., Tran, B., Chen, K., Sinclair, M.~D., and Venkataraman, S.
\newblock {PAL}: A variability-aware policy for scheduling {ML} workloads in
  {GPU} clusters.
\newblock In \emph{SC24: International Conference for High Performance
  Computing, Networking, Storage and Analysis}, pp.\  1--18. IEEE, 2024.
\newblock \doi{10.1109/SC41406.2024.00032}.

\bibitem[Kimpson et~al.(2024)Kimpson, Suvorova, Middleton, Liu, Melatos, Evans,
  and Moran]{kimpson2024anc}
Kimpson, T., Suvorova, S., Middleton, H., Liu, C., Melatos, A., Evans, R., and
  Moran, W.
\newblock Adaptive cancellation of mains power interference in continuous
  gravitational wave searches with a hidden {Markov} model.
\newblock \emph{Physical Review D}, 110:\penalty0 122004, 2024.
\newblock \doi{10.1103/PhysRevD.110.122004}.

\bibitem[Kimpson et~al.(2026)Kimpson, Baker, and Graham]{kimpson2026power}
Kimpson, T., Baker, M., and Graham, E.
\newblock Can power draw constrain covert compute? {L}imits of analogue
  verification for {AI} governance, 2026.
\newblock URL \url{https://arxiv.org/abs/2610.07476}.

\bibitem[Ko \& Zhu(2025)Ko and Zhu]{ko2025widearea}
Ko, M.-S. and Zhu, H.
\newblock Wide-area power system oscillations from large-scale {AI} workloads,
  August 2025.
\newblock URL \url{https://arxiv.org/abs/2508.16457}.

\bibitem[Kwon et~al.(2023)Kwon, Li, Zhuang, Sheng, Zheng, Yu, Gonzalez, Zhang,
  and Stoica]{kwon2023vllm}
Kwon, W., Li, Z., Zhuang, S., Sheng, Y., Zheng, L., Yu, C.~H., Gonzalez, J.~E.,
  Zhang, H., and Stoica, I.
\newblock Efficient memory management for large language model serving with
  {PagedAttention}.
\newblock In \emph{Proceedings of the 29th Symposium on Operating Systems
  Principles (SOSP '23)}, pp.\  611--626. ACM, 2023.
\newblock \doi{10.1145/3600006.3613165}.

\bibitem[Lin et~al.(2025)Lin, Jiang, Song, Zhao, Yu, Wang, Wang, Shi, Shi, Jia,
  Liu, Wang, Lin, Liu, Panda, and Li]{lin2025stragglers}
Lin, J., Jiang, Z., Song, Z., Zhao, S., Yu, M., Wang, Z., Wang, C., Shi, Z.,
  Shi, X., Jia, W., Liu, Z., Wang, S., Lin, H., Liu, X., Panda, A., and Li, J.
\newblock Understanding stragglers in large model training using what-if
  analysis.
\newblock In \emph{19th USENIX Symposium on Operating Systems Design and
  Implementation (OSDI 25)}, 2025.
\newblock URL \url{https://arxiv.org/abs/2505.05713}.

\bibitem[McCandlish et~al.(2018)McCandlish, Kaplan, Amodei, and {OpenAI Dota
  Team}]{mccandlish2018empirical}
McCandlish, S., Kaplan, J., Amodei, D., and {OpenAI Dota Team}.
\newblock An empirical model of large-batch training, 2018.
\newblock URL \url{https://arxiv.org/abs/1812.06162}.

\bibitem[O'Gara et~al.(2025)O'Gara, Kulp, Hodgkins, Petrie, Immler, Aysu, Basu,
  Bhasin, Picek, and Srivastava]{ogara2025hardware}
O'Gara, A., Kulp, G., Hodgkins, W., Petrie, J., Immler, V., Aysu, A., Basu, K.,
  Bhasin, S., Picek, S., and Srivastava, A.
\newblock Hardware-enabled mechanisms for verifying responsible {AI}
  development.
\newblock In \emph{{ICML} 2025 Workshop on Technical {AI} Governance}, 2025.
\newblock URL \url{https://arxiv.org/abs/2505.03742}.

\bibitem[Patel et~al.(2023)Patel, Choukse, Zhang, Goiri, Warrier, Mahalingam,
  and Bianchini]{patel2023polca}
Patel, P., Choukse, E., Zhang, C., Goiri, {\'I}., Warrier, B., Mahalingam, N.,
  and Bianchini, R.
\newblock {POLCA}: Power oversubscription in {LLM} cloud providers, 2023.
\newblock URL \url{https://arxiv.org/abs/2308.12908}.

\bibitem[Patel et~al.(2024{\natexlab{a}})Patel, Choukse, Zhang, Goiri, Warrier,
  Mahalingam, and Bianchini]{patel2024characterizing}
Patel, P., Choukse, E., Zhang, C., Goiri, {\'I}., Warrier, B., Mahalingam, N.,
  and Bianchini, R.
\newblock Characterizing power management opportunities for {LLMs} in the
  cloud.
\newblock In \emph{Proceedings of the 29th ACM International Conference on
  Architectural Support for Programming Languages and Operating Systems, Volume
  3 (ASPLOS '24)}, pp.\  207--222. ACM, 2024{\natexlab{a}}.
\newblock \doi{10.1145/3620666.3651329}.

\bibitem[Patel et~al.(2024{\natexlab{b}})Patel, Choukse, Zhang, Shah, Goiri,
  Maleki, and Bianchini]{patel2024splitwise}
Patel, P., Choukse, E., Zhang, C., Shah, A., Goiri, {\'I}., Maleki, S., and
  Bianchini, R.
\newblock Splitwise: Efficient generative {LLM} inference using phase
  splitting.
\newblock In \emph{2024 ACM/IEEE 51st Annual International Symposium on
  Computer Architecture (ISCA)}, pp.\  118--132, 2024{\natexlab{b}}.
\newblock \doi{10.1109/ISCA59077.2024.00019}.

\bibitem[Penedo et~al.(2024)Penedo, Kydl{\'i}{\v{c}}ek, Ben~Allal, Lozhkov,
  Mitchell, Raffel, Von~Werra, and Wolf]{penedo2024fineweb}
Penedo, G., Kydl{\'i}{\v{c}}ek, H., Ben~Allal, L., Lozhkov, A., Mitchell, M.,
  Raffel, C., Von~Werra, L., and Wolf, T.
\newblock The {FineWeb} datasets: Decanting the web for the finest text data at
  scale.
\newblock In \emph{Advances in Neural Information Processing Systems (Datasets
  and Benchmarks Track)}, 2024.
\newblock URL \url{https://arxiv.org/abs/2406.17557}.

\bibitem[Percival \& Walden(1993)Percival and Walden]{percival1993spectral}
Percival, D.~B. and Walden, A.~T.
\newblock \emph{Spectral Analysis for Physical Applications: Multitaper and
  Conventional Univariate Techniques}.
\newblock Cambridge University Press, 1993.
\newblock \doi{10.1017/CBO9780511622762}.

\bibitem[Rahman \& Tajdari(2026)Rahman and Tajdari]{rahman2026telemetry}
Rahman, R. and Tajdari, S.
\newblock Detecting hidden {ML} training with zero-overhead telemetry, 2026.
\newblock URL \url{https://arxiv.org/abs/2606.19262}.

\bibitem[Ramdas et~al.(2023)Ramdas, Gr{\"u}nwald, Vovk, and
  Shafer]{ramdas2023savi}
Ramdas, A., Gr{\"u}nwald, P., Vovk, V., and Shafer, G.
\newblock Game-theoretic statistics and safe anytime-valid inference.
\newblock \emph{Statistical Science}, 38\penalty0 (4):\penalty0 576--601, 2023.
\newblock \doi{10.1214/23-STS894}.

\bibitem[Sastry et~al.(2024)Sastry, Heim, Belfield, Anderljung, Brundage,
  Hazell, O'Keefe, Hadfield, Ngo, Pilz, Gor, Bluemke, Shoker, Egan, Trager,
  Avin, Weller, Bengio, and Coyle]{sastry2024computing}
Sastry, G., Heim, L., Belfield, H., Anderljung, M., Brundage, M., Hazell, J.,
  O'Keefe, C., Hadfield, G.~K., Ngo, R., Pilz, K., Gor, G., Bluemke, E.,
  Shoker, S., Egan, J., Trager, R.~F., Avin, S., Weller, A., Bengio, Y., and
  Coyle, D.
\newblock Computing power and the governance of artificial intelligence, 2024.
\newblock URL \url{https://arxiv.org/abs/2402.08797}.

\bibitem[Shavit(2023)]{shavit2023chinchilla}
Shavit, Y.
\newblock What does it take to catch a {Chinchilla}? {Verifying} rules on
  large-scale neural network training via compute monitoring, 2023.
\newblock URL \url{https://arxiv.org/abs/2303.11341}.

\bibitem[Shazeer et~al.(2017)Shazeer, Mirhoseini, Maziarz, Davis, Le, Hinton,
  and Dean]{shazeer2017moe}
Shazeer, N., Mirhoseini, A., Maziarz, K., Davis, A., Le, Q.~V., Hinton, G., and
  Dean, J.
\newblock Outrageously large neural networks: The sparsely-gated
  mixture-of-experts layer.
\newblock In \emph{5th International Conference on Learning Representations
  (ICLR 2017)}, 2017.
\newblock URL \url{https://arxiv.org/abs/1701.06538}.

\bibitem[Streit \& Barrett(1990)Streit and Barrett]{streit1990frequency}
Streit, R.~L. and Barrett, R.~F.
\newblock Frequency line tracking using hidden {Markov} models.
\newblock \emph{IEEE Transactions on Acoustics, Speech, and Signal Processing},
  38\penalty0 (4):\penalty0 586--598, 1990.
\newblock \doi{10.1109/29.52700}.

\bibitem[Suvorova et~al.(2016)Suvorova, Sun, Melatos, Moran, and
  Evans]{suvorova2016hmm}
Suvorova, S., Sun, L., Melatos, A., Moran, W., and Evans, R.~J.
\newblock Hidden {Markov} model tracking of continuous gravitational waves from
  a neutron star with wandering spin.
\newblock \emph{Physical Review D}, 93:\penalty0 123009, 2016.
\newblock \doi{10.1103/PhysRevD.93.123009}.

\bibitem[Thomson(1982)]{thomson1982multitaper}
Thomson, D.~J.
\newblock Spectrum estimation and harmonic analysis.
\newblock \emph{Proceedings of the IEEE}, 70\penalty0 (9):\penalty0 1055--1096,
  1982.
\newblock \doi{10.1109/PROC.1982.12433}.

\bibitem[Ville(1939)]{ville1939etude}
Ville, J.
\newblock \emph{{\'E}tude critique de la notion de collectif}.
\newblock Gauthier-Villars, Paris, 1939.

\bibitem[Vovk et~al.(2005)Vovk, Gammerman, and Shafer]{vovk2005algorithmic}
Vovk, V., Gammerman, A., and Shafer, G.
\newblock \emph{Algorithmic Learning in a Random World}.
\newblock Springer, 2005.

\bibitem[Wu et~al.(2024)Wu, Wang, Yu, Yang, Wu, Duan, Yang, Wang, Qu, and
  Zhang]{wu2024falcon}
Wu, T., Wang, W., Yu, Y., Yang, S., Wu, W., Duan, Q., Yang, G., Wang, J., Qu,
  L., and Zhang, L.
\newblock {FALCON}: Pinpointing and mitigating stragglers for large-scale
  hybrid-parallel training, 2024.
\newblock URL \url{https://arxiv.org/abs/2410.12588}.

\bibitem[Yu et~al.(2022)Yu, Jeong, Kim, Kim, and Chun]{yu2022orca}
Yu, G.-I., Jeong, J.~S., Kim, G.-W., Kim, S., and Chun, B.-G.
\newblock {Orca}: A distributed serving system for transformer-based generative
  models.
\newblock In \emph{16th USENIX Symposium on Operating Systems Design and
  Implementation (OSDI 22)}, pp.\  521--538. USENIX Association, 2022.
\newblock URL \url{https://www.usenix.org/conference/osdi22/presentation/yu}.

\end{thebibliography}
